\documentclass[12pt,amsmath,amssymb]{article}
\pdfoutput=1
\usepackage{epsfig,amsmath,amssymb} 
\usepackage{graphicx}
\usepackage{dcolumn}
\usepackage{bm}
\usepackage{hyperref}
\usepackage{xcolor}
\hypersetup{
    colorlinks=true,       
    linkcolor=black,          
    citecolor=blue,        
    filecolor=magenta,      
    urlcolor=blue           
}
\usepackage{cite}
\renewcommand\theequation{\arabic{section}.\arabic{equation}}
\newcommand{\be}{\begin{equation} } 
\newcommand{\ee}{\end{equation} } 
\newcommand{\ba}{\begin{array} } 
\newcommand{\ea}{\end{array} } 
\newcommand{\bear}{\begin{eqnarray} } 
\newcommand{\eear}{\end{eqnarray} } 
\newcommand{\met}{ E_T\!\!\! \!\! \! \slash \;\, } 
\newcommand{\Tr}{{\rm Tr}}

\title{
\vspace*{-2.3cm}  
\begin{flushright}
\normalsize{ \small Fermilab-PUB-17-555-T
  }
\end{flushright}
\vspace{1.5cm}
\Large  
\textbf{Collider Tests of the Renormalizable Coloron Model} \\ \vspace{1cm}
}

\author{{\bf Yang Bai}$^\star$ and {\bf Bogdan A. Dobrescu}$^\diamond$
\vspace{5mm}
\\
$^\star$\normalsize\emph{Department of Physics, University of Wisconsin-Madison, Madison, WI 53706, USA}  \vspace{1mm} \\
$^\diamond$\normalsize\emph{Theoretical Physics Department, Fermilab, Batavia, IL 60510, USA}
}
\date{September 7, 2018}

\begin{document}  
\setcounter{page}{0}  
\maketitle  

\vspace*{1cm}  
\begin{abstract} 
The coloron, a massive version of the gluon present in gauge extensions of QCD, has been searched for at the LHC as a dijet or top quark pair resonance. We point out that in the Renormalizable Coloron Model (ReCoM) with a minimal field content to break the gauge symmetry, a color-octet scalar and a singlet scalar are naturally lighter than the coloron because they are pseudo Nambu-Goldstone bosons. Consequently, the coloron may predominantly decay into scalar pairs, leading to novel signatures at the LHC. When the color-octet scalar is lighter than the singlet, or when the singlet mass is above roughly 1 TeV, the signatures consist of multi-jet resonances of multiplicity up to 12, including topologies with multi-prong jet substructure, slightly displaced vertices, and sometimes a top quark pair. When the singlet is the lightest ReCoM boson and lighter than about 1 TeV, its main decays ($W^+W^-$, $\gamma Z$, $ZZ$) arise at three loops. The LHC signatures then involve two or four boosted electroweak bosons, often originating from highly displaced vertices, plus one or two pairs of prompt jets or top quarks. 
\end{abstract} 
  
\thispagestyle{empty}  
\newpage  
  
\setcounter{page}{1}  
  
\hypersetup{linktocpage} 
\tableofcontents 
\hypersetup{linkcolor=red} 

\bigskip

\section{Introduction}
\label{sec:intro}

Massive elementary particles of  spin-1 are well-behaved at high-energies only if they are associated with a spontaneously broken gauge symmetry \cite{Cheng:1985bj}.
The sector responsible for breaking the gauge symmetry necessarily includes additional bosons (Higgs-like scalars or composite $\rho$-like fields)
that couple to the gauge boson, and are not much heavier (but could be much lighter) than it. 
As a result, the additional bosons may drastically affect the phenomenology of a new massive gauge boson (see, {\it e.g.}, \cite{Bai:2010dj,Dobrescu:2013gza}).

The coloron \cite{Hill:1991at,Hill:1993hs,Chivukula:1996yr,Simmons:1996fz} is a color-octet spin-1 particle that is present in gauge extensions of the QCD. The classic example
\cite{Hall:1985wz,Frampton:1987dn,Bagger:1987fz} is an  $SU(3)_1\times SU(3)_2$ gauge symmetry broken down to the $SU(3)_c$ gauge group of QCD.
If all SM quarks are triplets under one of the $SU(3)$ groups, then the coloron has flavor-universal couplings \cite{Chivukula:1996yr, Simmons:1996fz}, and 
dijet or top quark pair resonance searches at the LHC are sensitive to $s$-channel production of the coloron. 

The spontaneous breaking of  $SU(3)_1\times SU(3)_2$  is usually assumed to be due to a bi-fundamental scalar field that acquires a diagonal vacuum expectation value (VEV).
A scalar potential that achieves that was considered in \cite{Chivukula:1996yr, Simmons:1996fz}, and involved a mass term and two quartic terms.
It was pointed out  in  \cite{Bai:2010dj} that an additional term, of dimension 3, in the potential is consistent 
with the gauge and Lorentz symmetries. The presence of that dimension-3 term is in fact useful, as it leads to a mass 
for a Nambu-Goldstone boson that would otherwise be phenomenologically problematic. The regions of parameter space where the desired 
symmetry breaking pattern is achieved have been derived in \cite{Bai:2017zhj}.
We refer to the flavor-universal  $SU(3)_1\times SU(3)_2$ theory, with the most general terms of dimension up to four in the potential, as 
the Renormalizable Coloron Model (ReCoM).

Even though the ReCoM is a simple extension of the Standard Model (SM), its phenomenology is rich and has not been fully explored thus far.
Besides the coloron, there are three physical spin-0 particles \cite{Bai:2010dj}: a color-octet, $\Theta$, and two singlets, $\phi_I$ and $\phi_R$.
The constraints on the latter, which is a radial mode, have been derived in  \cite{Chivukula:2013xka,Chivukula:2014rka}
based on Higgs measurements, direct searches at the LHC,  
and electroweak observables. 
The processes discussed in Ref.~\cite{Bai:2010dj} rely on the existence of a vector-like quark, which is not part of the minimal structure of the ReCoM.

In this paper, we systematically study the novel collider signatures in the ReCoM and point out new search strategies for discovering the coloron and the 
associated scalars. 
The color-octet scalar and $\phi_I$ are pseudo-Nambu-Goldstone bosons (pNGB's), and thus are naturally lighter than the coloron. 
Depending on the scalar masses and on the amount of mixing of the $SU(3)_1\times SU(3)_2$ gauge bosons, the main 
decay channel of the coloron could be either $G^\prime_\mu \rightarrow \Theta \phi_I$ or $\Theta \Theta$, while the dijet branching fraction may be 
below 10\%.
 
The $\Theta$ and $\phi_I$ scalars have competing multi-body or multi-loop decays, as well as cascade decays in association with SM quark-antiquark pairs. 
As a result, the coloron appears as an $s$-channel multi-jet resonance, with two or more jet sub-clusters, and a total jet multiplicity as high as 12  when there are no top quarks produced.
Channels that include $t\bar t$ pairs have even more complicated final states.
For a lighter $\phi_I$, below around 1 TeV, its main decay channels are into two electroweak gauge bosons. The collider signatures in that case include boosted
$W^+W^-$, $\gamma Z$ or $ZZ$ resonances, which typically originate from displaced vertices, plus a number of prompt quark jets or $t\bar t$ pairs.  
Currently, there are no dedicated searches at the LHC for these classes of signatures. 
The ReCoM thus provides a motivation for new searches at the LHC with complicated hadronic final states, various boosted systems, and displaced vertices.

The paper is organized as follows. We first work out the properties of the scalars in the ReCoM including their 
 interactions and branching fractions in Section~\ref{sec:model}. In Section~\ref{sec:production}, we calculate the production cross sections in phenomenologically relevant channels. In Section~\ref{sec:LHC}, we identify the LHC signatures for different mass spectra and comment on search strategies. Our conclusions and two tables that summarize the main collider signatures are included 
 in Section~\ref{sec:conclusions}. In Appendix A, we derive the constraints on the scalar masses imposed by the condition that the color-conserving vacuum is the global minimum of the potential.
 In Appendix B we outline the computation of the partial widths for coloron decays into scalars.

\bigskip

\section{Renormalizable $SU(3)_1 \times SU(3)_2$ model}
\label{sec:model}
\setcounter{equation}{0}

We study a gauge extension of the SM with an  $SU(3)_1 \times SU(3)_2 \times SU(2)_W \times U(1)_Y$ gauge group and a new scalar 
field $\Sigma$ transforming in the $(3, \bar{3}, 1,0)$ representation.  The most general renormalizable potential that involves only $\Sigma$ has four terms:
\be
V(\Sigma) \,=\, -m_\Sigma^2\, \Tr(\Sigma \Sigma^\dagger) \, -\,  \left( \mu_\Sigma  \; {\rm det} \, \Sigma + {\rm H.c.} \right)  +
\frac{\lambda}{2}\left[ \Tr \left(\Sigma \Sigma^\dagger\right)  \right]^2 \,+\,\frac{\kappa}{2}\,\Tr \left(\Sigma \Sigma^\dagger \Sigma \Sigma^\dagger \right) ~.
\label{eq:sigma-pot}
\ee
The squared-mass parameter $m_\Sigma^2$ can be positive or negative. The phase-rotation freedom of $\Sigma$ allows us to choose the mass parameter 
$\mu_\Sigma$ to be real and positive  without loss of generality.  
At least one of the dimensionless couplings  $\lambda$ and $\kappa$ must be positive. If one of them is negative, then the potential is bounded from below only when 
both $\lambda + \kappa$ and $3\lambda + \kappa$ are positive \cite{Bai:2017zhj}.

At the renormalizable level, there is also a potential term for $\Sigma$ coupling to the SM Higgs doublet: $\mbox{Tr}(\Sigma \Sigma^\dagger) \, H H^\dagger$, which modifies the Higgs boson properties~\cite{Chivukula:2013xka,Chivukula:2014rka}. In the following, we will assume a small coupling for this interaction so that the modifications to the Higgs boson properties are below the current experimental sensitivity. 

\subsection{Bosonic mass spectrum}

As studied in detail in Ref.~\cite{Bai:2017zhj} (see also \cite{Nardi:2011st,Espinosa:2012uu}), and depending on the values of the four parameters in Eq.~(\ref{eq:sigma-pot}), there are three possible vacua with the gauge symmetries given by the diagonal subgroup $SU(3)_c$, or $SU(2)_1 \times SU(2)_2 \times U(1)$, or the full $SU(3)_1 \times SU(3)_2$. 
In ReCoM the phenomenologically viable regions of parameter space are those where the global minimum has the $SU(3)_c$ symmetry, identified with the QCD gauge group. These regions of parameter space are summarized in Appendix A. 

The $\Sigma$ field  is a $3\times 3$ matrix with complex entries, and its VEV is 
\be
\langle \Sigma \rangle \, = \, \frac{f_{\Sigma}}{\sqrt{6}}\,  \mathbb{I}_3 ~~~,
\ee
with $\mathbb{I}_3 $ the unit matrix, and the scale of $SU(3)_1 \times SU(3)_2$ breaking given by 
\be
f_\Sigma  = \frac{ \sqrt{3}}{ \sqrt{2} \, (3 \lambda + \kappa)} \left(
\sqrt{4(3 \,\lambda + \kappa) \, m^2_\Sigma + \mu^2_\Sigma} + \mu_\Sigma\right)  > 0  ~~~.
\label{eq:fsigma}
\ee

Eight of the degrees of freedom in $\Sigma$ become the longitudinal degrees of freedom of the heavy gauge boson (the coloron), while the remaining 
ten degrees of freedom form a color-octet real scalar, $\Theta^a$, and two singlet real scalars, $\phi_R$ and $\phi_I$. The relation between these physical states and the 
uneaten degrees of freedom in $\Sigma$ is
\be
 \Sigma =  \frac{f_{\Sigma} +  \phi_R + i \, \phi_I}{\sqrt{6}} \, \mathbb{I}_3 + \Theta^a \, T^a    ~~~,
 \label{eq:Sigma-parametrization}
\ee
where $T^a$ are the $SU(3)_c$ generators, normalized as $\mbox{Tr}(T^a T^b) = \frac{1}{2}\delta^{ab}$. Since the new scalars  in the ReCoM do not have tree-level 
couplings to fermions, they are $\mathcal{P}$-even. Under charge conjugation the scalars transform as $\Sigma \xrightarrow{\mathcal{C}} \Sigma^*$, which implies $\Theta^a T^a \xrightarrow{\mathcal{C}} \Theta^a (T^a)^*$, $\phi_R \xrightarrow{\mathcal{C}} \phi_R$, and $\phi_I \xrightarrow{\mathcal{C}} - \phi_I$. Thus, $\phi_I$ is a $\mathcal{C}\mathcal{P}$-odd scalar which is also $\mathcal{C}$-odd. 
Its $\mathcal{C}\mathcal{P}$ property will determine its couplings to SM particles and lifetime, which will be discussed later. The squared-masses of the three scalar particles are given by
\bear
&& M^2_\Theta = \frac{\kappa}{3}\,f_\Sigma^2  + \sqrt{\frac{2}{3}}\; \mu_\Sigma\,f_\Sigma ~~ \,, 
\nonumber \\ [2mm]
&& M^2_{\phi_I} = \sqrt{\frac{3}{2}}\;\mu_\Sigma\,f_\Sigma ~~\,,
\nonumber \\ [2mm]
&& M^2_{\phi_R} = \left( \lambda + \frac{\kappa}{3}\right)\,f_\Sigma^2 - \frac{\mu_\Sigma}{\sqrt{6}}\; f_\Sigma  ~~ \,.
\label{eq:scalar-masses}
\eear

Imposing that the $SU(3)_c$-preserving vacuum is the global minimum of the potential, we prove in Appendix A that there is an upper bound for 
the ratio of the $\phi_I$ and $\Theta$ masses: $M_{\phi_I} / M_\Theta < 2.1$.
In the limit of $\mu_\Sigma \ll f_\Sigma$, we have $M_{\phi_I} \ll M_{\phi_R}, M_\Theta$, indicating that the  color-singlet scalar $\phi_I$ becomes the light pNGB 
associated with the spontaneous breaking of a global $U(1)_\Sigma$ symmetry acting on $\Sigma$. The other color-singlet scalar $\phi_R$ has a mass of 
order $\sqrt{\lambda + \kappa/3} \, f_\Sigma$, unless a 
fine-tuning of parameters makes it much lighter. 

The limit of $\kappa \ll 1$ and  $\mu_\Sigma \ll f_\Sigma$ is also interesting, as it enhances the global symmetry of the potential to $SO(18)$, which is the symmetry that rotates the degrees of freedom of $\Sigma$. This symmetry is spontaneously broken 
down to $SO(17)$ by $\langle \Sigma \rangle$, so that there are 17 light pNGB's in this case, which are the degrees of freedom within the longitudinal coloron,  the color-octet scalar $\Theta$, and $\phi_I$. 
In that limit both $\Theta$ and $\phi_I$ are naturally lighter than the coloron and $\phi_R$.

Let us label the  $SU(3)_1 \times SU(3)_2$ gauge bosons by $G_1$ and $G_2$, and the corresponding gauge couplings by $h_1$ and $h_2$.
One linear combination of the gauge bosons becomes the massless QCD gluon ($G$)  and the orthogonal combination is massive, the coloron ($G'$):
\bear
&& G^\mu = G^\mu_1 \, \cos{\theta} + G^\mu_2 \, \sin{\theta}   ~~ \,, 
\nonumber \\ [2mm] 
&& G^{\prime\mu} =  G^\mu_1 \, \sin{\theta} - G^\mu_2 \, \cos{\theta}  ~~~,
\eear
where the mixing angle satisfies
\be
\tan{\theta} = \frac{h_1}{h_2}  ~~~.
\label{eq:tan-theta}
\ee
 The QCD gauge coupling is related to the $SU(3)_1 \times SU(3)_2$ couplings by 
 \be
 g_s = \frac{h_1 h_2}{ \sqrt{h_1^2 + h_2^2}}   ~~,
 \label{eq:gs}
 \ee
and the coloron mass is 
\be
M_{G^\prime} = \frac{g_s}{\sqrt{6}} \, \left( \tan{\theta} + \frac{1}{\tan\theta}  \right)  f_\Sigma ~~. 
\label{eq:Gprime-mass}
\ee

The gauge couplings are perturbative provided $3 h_i^2 / (16 \pi^2) \lesssim 1$,  based on naive dimensional analysis.
These upper limits together with Eq.~(\ref{eq:gs})  translate into a constraint on $\tan{\theta}$:  $\, 0.15\lesssim \tan{\theta} \lesssim 6.7$. 
We will use this range of $\tan{\theta}$ for our phenomenological analysis. We point out, however, that 
the tree-level analysis is more under control if the upper limit on the 
gauge couplings is lowered, {\it e.g},
$3 h_i^2 / (16 \pi^2) \lesssim 1/3$, which gives $0.25\lesssim \tan{\theta} \lesssim 4$. 

\subsection{Interactions of the new bosons}

The interactions among physical scalars include a trilinear $\Theta$ term,
\be
\overline\mu_\Sigma  \, d^{abc}  \Theta^a \Theta^b \Theta^c\,     ~~,
\label{eq:trilinearTheta}
\ee
where  $d^{abc} $ is the totally-symmetric color tensor.
The coefficient $\overline\mu_\Sigma$ has mass dimension one, and receives contributions from the trilinear term in the $V(\Sigma)$ potential
as well as from a quartic term in $V(\Sigma)$ with one insertion of the $\Sigma$ VEV:
\bear
\overline\mu_\Sigma  &=& \sqrt{\frac{3}{2}} \,\kappa\,f_\Sigma - \mu_\Sigma  \nonumber \\
&=& \frac{3\,g_s \,(1+\tan^2\!{\theta})}{2\,M_{G'}\,\tan\!{\theta}}\,\left( M_\Theta^2 - \frac{8}{9}\,M^2_{\phi_I}\right) ~~~,
\label{eq:mubar}
  \eear
where in the second line we have used the relations between parameters and boson masses from Eqs.~(\ref{eq:scalar-masses}) and (\ref{eq:Gprime-mass}). 
There are also trilinear interactions involving one $\phi_R$ singlet and  either two color-octets or two $\phi_I$ singlets,
\be
  \frac{1}{2}\,\left[ (\lambda + \kappa)\,f_\Sigma + \frac{\mu_\Sigma}{\sqrt{6}}  \right]\phi_R\, \Theta^a \Theta^a  + 
 \frac{1}{2}\,\left[ \left( \lambda + \frac{\kappa}{3} \right) \, f_\Sigma + \frac{2\,\mu_\Sigma}{\sqrt{6}}  \right] \phi_R \, \phi_I^2 ~~~.
\label{eq:3scalar}
\ee
The only other trilinear  scalar interaction involves three $\phi_R$ and is less  phenomenologically important.

The interactions of a single coloron with scalars are given by
\be
\frac{g_s}{\tan \theta} \, G_\mu^{\prime\,a} \left[ \frac{1}{\sqrt{6}}  \left(  1 + \tan^2\!\theta \right)  \, \left( \phi_I \partial^\mu \Theta^a  - \Theta^a \partial^\mu\phi_I \right)   
 - \frac{1}{2} \left(  1 - \tan^2\!\theta \right)   \,f^{abc}\,  \Theta^b \partial^\mu\,\Theta^c  \right] ~~,
\label{eq:GpScalars}
\ee
where $f^{abc} $ is the totally-antisymmetric color tensor. Note that, for $\tan{\theta} =1$, the coupling of $G'_\mu$ to two $\Theta$'s vanishes. This is because when 
$h_1=h_2$ [see Eq.~\eqref{eq:tan-theta}] there is an interchanging $\mathcal{Z}_2$ symmetry, under which $G^\mu_1 \leftrightarrow G^\mu_2$, $G_\mu \leftrightarrow G_\mu$, $G^\prime_\mu \leftrightarrow - G^\prime_\mu$, $\Sigma \rightarrow \Sigma^\dagger$, $\phi_I \rightarrow - \phi_I$ and $\Theta^a \rightarrow \Theta^a$.  

The couplings of a pair of colorons to a single scalar are 
\be
\frac{M^2_{G^\prime}}{f_\Sigma} \left( \sqrt{\frac{3}{2}}\,d^{abc} G_\mu^{\prime a} \, G^{\prime \mu b}\,\Theta^c \,+\, G_\mu^{\prime a} \, G^{\prime \mu a}\,\phi_R \right) \,.
\label{eq:GGT}
\ee
We will not display here the interactions of two colorons with two scalars, the quartic scalar interactions, or the coloron self-interactions (the latter are important if one of the $h_i$ gauge couplings is nearly nonperturbative).

For coloron interactions with fermions, we choose the simplest and anomaly-free model with the SM quarks as triplets under $SU(3)_1$ and singlets under $SU(3)_2$. The  coloron interactions with quarks are flavor-blind and have the same form as those of the gluon, with an extra factor of $\tan\theta$:
\be
g_s \tan{\theta}\,\bar{q} \gamma^\mu T^a G^{\prime a}_\mu \, q   ~~~. 
\ee

\subsection{Partial widths of the color-octet bosons}
\label{sec:ThetaWidth}

At tree-level and for $M_\Theta < M_{G^\prime}$, the main decay of the  color-octet scalar is into two  quark-antiquark pairs, via two off-shell colorons: $\Theta \to G^{\prime *}  G^{\prime *}   \to q \bar q  q' \bar q'$.
Using  Eqs.~(\ref{eq:GGT}) and (\ref{eq:Gprime-mass}), the 4-body width to leading order in $M_\Theta^2 \ll M_{G'}^2$ is 
parametrically of the order of 
\be
\Gamma \left( \Theta \to  q \bar q  q^\prime \bar q^\prime \right)  \sim \frac{\alpha_s^3} {\pi^2} \tan^2\!\theta(1+ \tan^2\!\theta )^2  \frac{M_\Theta^7}{M_{G'}^6}   ~~,
\ee
where the width is summed over the quark flavors, and $\alpha_s$ is the QCD coupling constant.
We have assumed here that the phase-space suppression is of the order of $(4\pi)^5$.
For $M_\Theta \sim 1$ TeV,  $M_{G'} \sim 4$ TeV and $\tan\theta \sim 0.3$, the 4-body width is at the keV scale. For larger $M_\Theta$, of order $M_{G'}/2$,
$\Gamma \left( \Theta \to  q \bar q  q' \bar q' \right) $ is larger by a few orders of magnitude.

At one loop, $\Theta$ decays into two gluons. The amplitude for this process gets two contributions. The first one is from diagrams with $\Theta$ running in the loop and a vertex involving the trilinear interaction (\ref{eq:trilinearTheta}). The second contribution is from diagrams with the coloron running in the loop and the $G'G' \Theta$ vertex given in Eq.~(\ref{eq:GGT}). Adding the two contributions, we have the following effective interaction:
\be
\frac{C_\Theta}{f_\Sigma} 
\left(  \frac{g_s^2}{16\pi^2} \right) \, d^{abc}\, \Theta^a \,  G^b_{\alpha\beta} G^{c\,\alpha\beta} ~~,
\label{eq:effectiveThetaGluons}
\ee
where the dimensionless coefficient is 
\be
C_\Theta = 6\sqrt{2}\,\left( \frac{\pi^2}{9} - 1\right)
 \frac{\overline{\mu}_\Sigma \,f_\Sigma }{M_\Theta^2} + \frac{3\sqrt{6}}{32}\,\mathcal{A}\left[M^2_\Theta/(4\,M^2_{G'}) \right]~~.
\label{eq:theta-twogluon}
\ee
Here, the function $\mathcal{A}(\tau) = [2\tau^2 + 3\tau + 3(2\tau -1)\arcsin^2\!{\sqrt{\tau}} \, ]\,\tau^{-2}$ for $\tau \leq 1$, which is similar to the $W^\pm$ loop contributions to the effective interaction 
of the Higgs boson with two photons. In the $\tau \rightarrow 0$ limit, the function $\mathcal{A}(\tau)$ reaches its minimum, $\mathcal{A}(0_+) \rightarrow 7$; $\mathcal{A}(\tau)$ increases 
monotonically with $\tau$, and $\mathcal{A}(1/4) \approx 7.42$. 
The ratio of the second term to the first term in Eq.~(\ref{eq:theta-twogluon}) has a simple formula:
\bear
r_{\mathcal{A}}  & = & \frac{3}{32\sqrt{2}\,(\pi^2 - 9)}\, \frac{\mathcal{A}\left[M^2_\Theta/(4\,M^2_{G'})\right]}{1 - 8 M_{\phi_I}^2/ (9 M_\Theta^2)} 
\nonumber \\ [2mm]
&\approx & 0.533 \left(  1 + \frac{8 M_{\phi_I}^2}{9 M_\Theta^2} +   \frac{11 M_\Theta^2}{210 M_{G'}^2} + \cdots   
\right)   
~~,
\label{eq:r_A}
\eear
where the ellipsis in the second line refers to higher-order terms in  $M_{\phi_I}^2 / M_\Theta^2$ and $M_\Theta^2/M_{G'}^2$.
The above expression shows that the two contributions to $\Theta \to gg$  are  comparable. 
Using the result for the scalar contribution given in  \cite{Bai:2010dj}, and including the interference with the coloron contribution, we find the $\Theta \to gg$ width:
\bear
\Gamma (\Theta \to gg) &=& \frac{15 \,\alpha_s^2 \; \overline\mu_\Sigma^2 }{128 \,\pi^3 \,M_{\Theta}}\,\left(\frac{\pi^2}{9} -1 \right)^{\! 2} \,\left(1 + r_{\mathcal{A}} \right)^2 
 \nonumber \\ [2mm]
&\approx &\frac{135 \,\alpha_s^2}{256 \,\pi^3}\,\left(\frac{\pi^2}{9} -1 \right)^{\! 2}\,\left(1 + r_{\mathcal{A}} \right)^2   \,\kappa\,M_{\Theta}\; ~, 
\label{eq:decaygg}
\eear
where the approximation in the second line is taken in the limit of $\mu_\Sigma \ll \kappa\,f_\Sigma$ (which implies $M_{\phi_I} \ll M_{\Theta}$). 
Note the accidental suppression of the above width by a factor of  $(\pi^2/9 -1)^2 \approx 9.3 \times 10^{-3}$.
The interference of the scalar and coloron contributions is constructive for  $ M_{\phi_I} <  6 \sqrt{2} \,  M_\Theta$, which is automatically satisfied in the ReCoM (see Appendix A).
For $M_\Theta = 1$~TeV,  $M_{G'} = 4$~TeV, $M_{\phi_I}=300$~GeV and $\tan\theta \sim 0.3$, the 1-loop 2-body width of $\Theta$ is around 2 MeV, much above the tree-level 4-body one.

\begin{figure}[t]
\begin{center}
\includegraphics[width=0.46\textwidth]{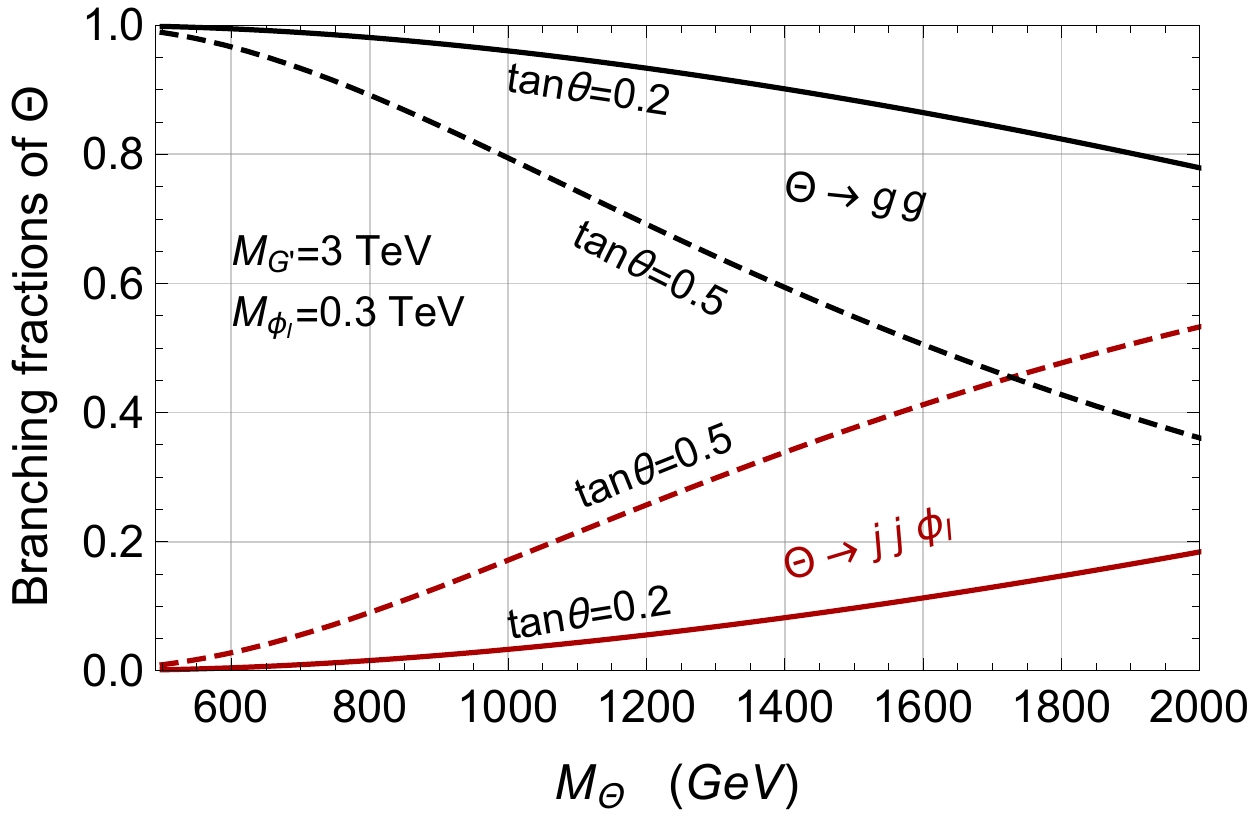} \hspace{3mm}
\includegraphics[width=0.46\textwidth]{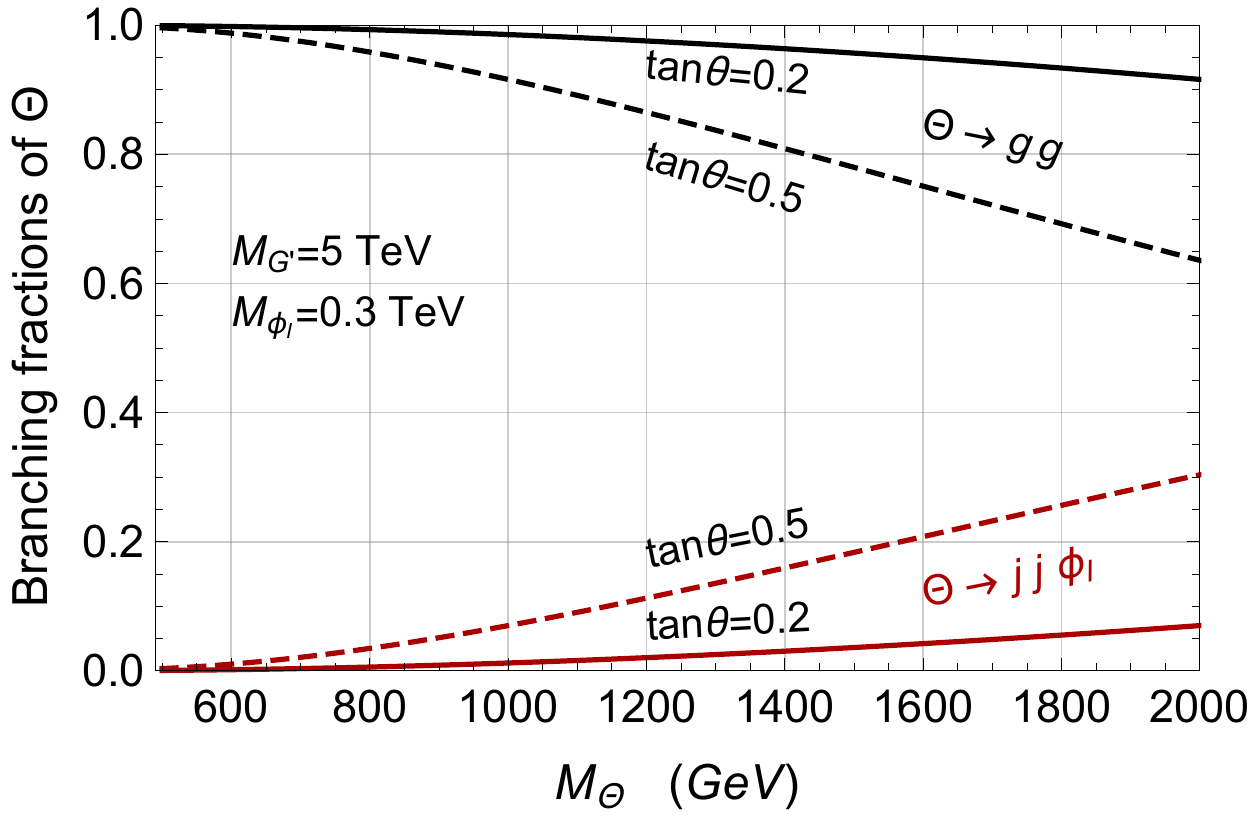}
\caption{Branching fractions of the color-octet scalar $\Theta$ as a function of its mass. 
For $\Theta \to j j \phi_I$ the width is summed over the five light quark flavors in the final state. 
The branching fraction for $\Theta \to t \bar t \phi_I$ (not shown) is slightly less than 1/5 of that for $\Theta \to j j \phi_I$. 
The coloron mass used here is  $M_{G'} = 3$ TeV (left panel) or 5 TeV (right panel), and the neutral scalar mass is 
fixed at $M_{\phi_I} = 300$ GeV.
The parameter $\tan{\theta}$, which determines the strength of the coloron couplings, is taken to be 0.2 (solid lines) or 0.5 (dashed lines).
 }
\label{fig:theta-branching}
\end{center}
\end{figure}

We also note that there is an effective interaction of $\Theta^a$ with two quarks, which can be generated at one loop, with a $G^\prime$ and a quark running in the loop. However, this scalar-fermion interaction is suppressed by the ratio of fermion mass over coloron mass. Even for the top quark, the estimated width is smaller by two orders of magnitude than the one of $\Theta \rightarrow gg$. 
We will ignore this small branching fraction in our analysis. 

If $M_\Theta > M_{\phi_I}$, then $\Theta$ also has 3-body  tree-level decays into a $\phi_I$ scalar and a quark-antiquark pair of same flavor, $\Theta \rightarrow \bar{q} q \,\phi_I$, through an off-shell $G'$. 
Neglecting the quark masses, the partial widths of $\Theta$ into a $\phi_I$ and either a pair of quark jets ($jj$) or $t \bar t$ satisfy
\be
 \hspace*{-.4cm}  \Gamma(\Theta \rightarrow jj \,\phi_I) \simeq 5\,\Gamma(\Theta \rightarrow \bar{t} t \,\phi_I) 
 \simeq \frac{5\,\alpha_s^2 M^5_{\Theta}}{576\,\pi M^4_{G^\prime}}  \!
  \left(  1 + \tan^2\!\theta \right)^2 
 {\cal F} \! \left(M_{\Theta }^2 / M_{G^\prime}^2 , M_{\phi_I}^2 /M_{G^\prime}^2\right) ,
   \label{eq:3bodyTheta}
\ee
where we defined the function
\bear
&&   \hspace*{-1.cm} 
{\cal F}(x,y) = \frac{12}{x^4} \left[ (y  \!- \! x) \left(1 - \frac{3}{2} (x  \!+ \! y) + \frac{1}{3} (x  \!- \! y)^2  \right) + 
 \frac{1}{2} \left(1 - 2(x  \!+ \! y) + x^2+ y^2    \rule{0mm}{5mm} \right) \log \!\left(\frac{x}{y } \right) 
 \right.
\nonumber \\  [2mm]
&&     \hspace*{-.7cm} 
\left. + \left(1  \!- \! x  \!- \! y \right)   \left(1  \!- \! 2 (x  \!+ \! y) + (x \! -  \!y)^2   \rule{0mm}{5mm}   \right)^{1/2}  
  \tanh^{-1}\!\!
  \left( \!\frac{  \left(1 - 2 (x \!+\!y) + (x\!-\!y)^2    \rule{0mm}{3.7mm}   \right)^{\! 1/2} }{ x - y - (x \!+ \!y)/(x \!- \!y)}\right)  
\!\right] ~.
   \label{eq:3bodyFunc}
\eear
 In Figure~\ref{fig:theta-branching}, we show  the branching fractions of 2- and 3-body decays as a function of $M_\Theta$ for fixed masses of $\phi_I$ and the coloron. 
Note that the 3-body branching fraction increases as $M_\Theta$ increases. For a fixed value of $M_\Theta$, increasing the mixing parameter $\tan{\theta}$ leads to an increase of the 3-body branching fraction. 
It is helpful to take the $M_\Theta, M_{\phi_I}\ll M_{G^\prime}$ limit (for any $M_\Theta > M_{\phi_I}$) in Eq.~(\ref{eq:3bodyTheta}), which corresponds to the following limit:
\be
  {\cal F}(x,y)    \underset{x,y \ll 1 \; \;\; \;\; }{ \longrightarrow} \!\!
  1 - 8 \frac{y}{x}   + 12 \,  \frac{y^2}{x^2} \, \log{\left( \frac{x}{y} \right)} 
 + 8 \, \frac{y^3}{x^3} -  \frac{y^4}{x^4}    ~~~.
\ee

\begin{figure}[t]
\begin{center}
\includegraphics[width=0.495\textwidth]{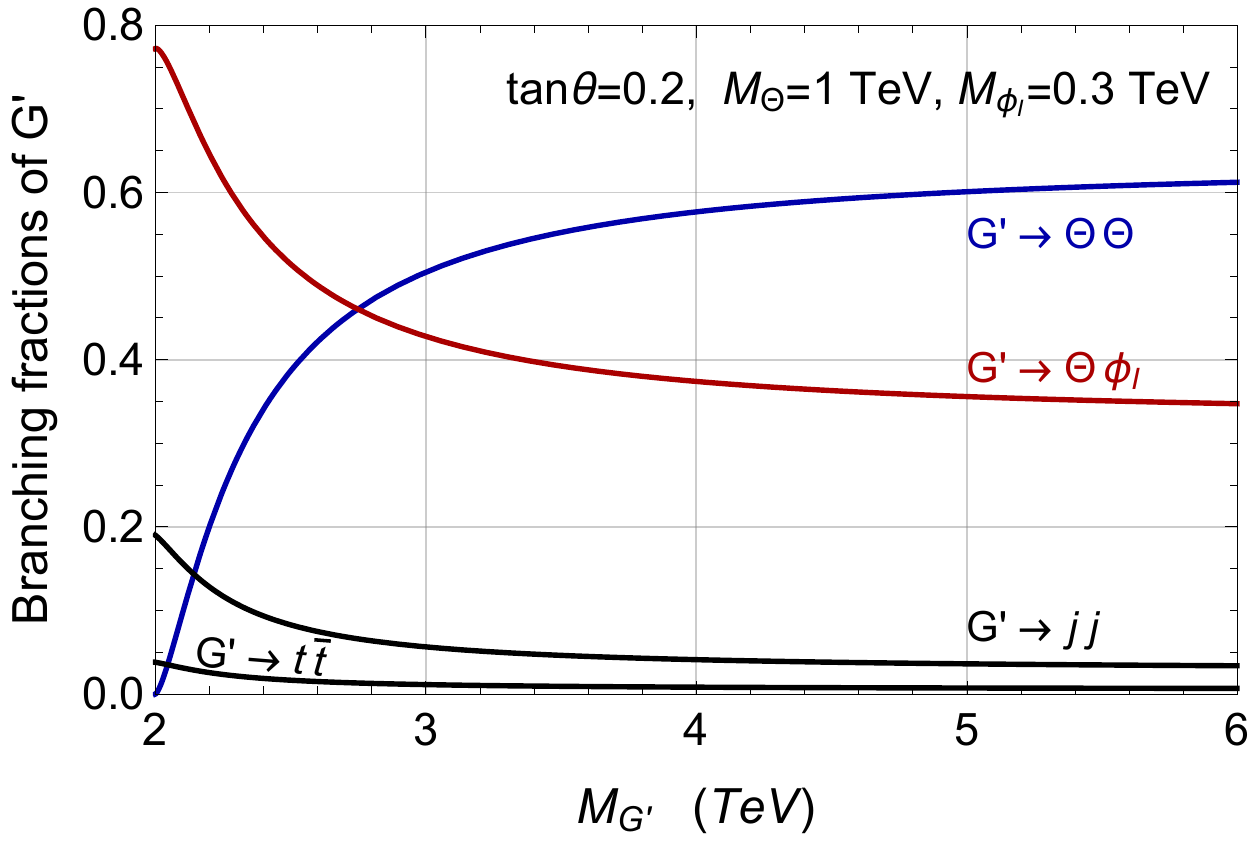} \hspace{-1.1mm}
\includegraphics[width=0.495\textwidth]{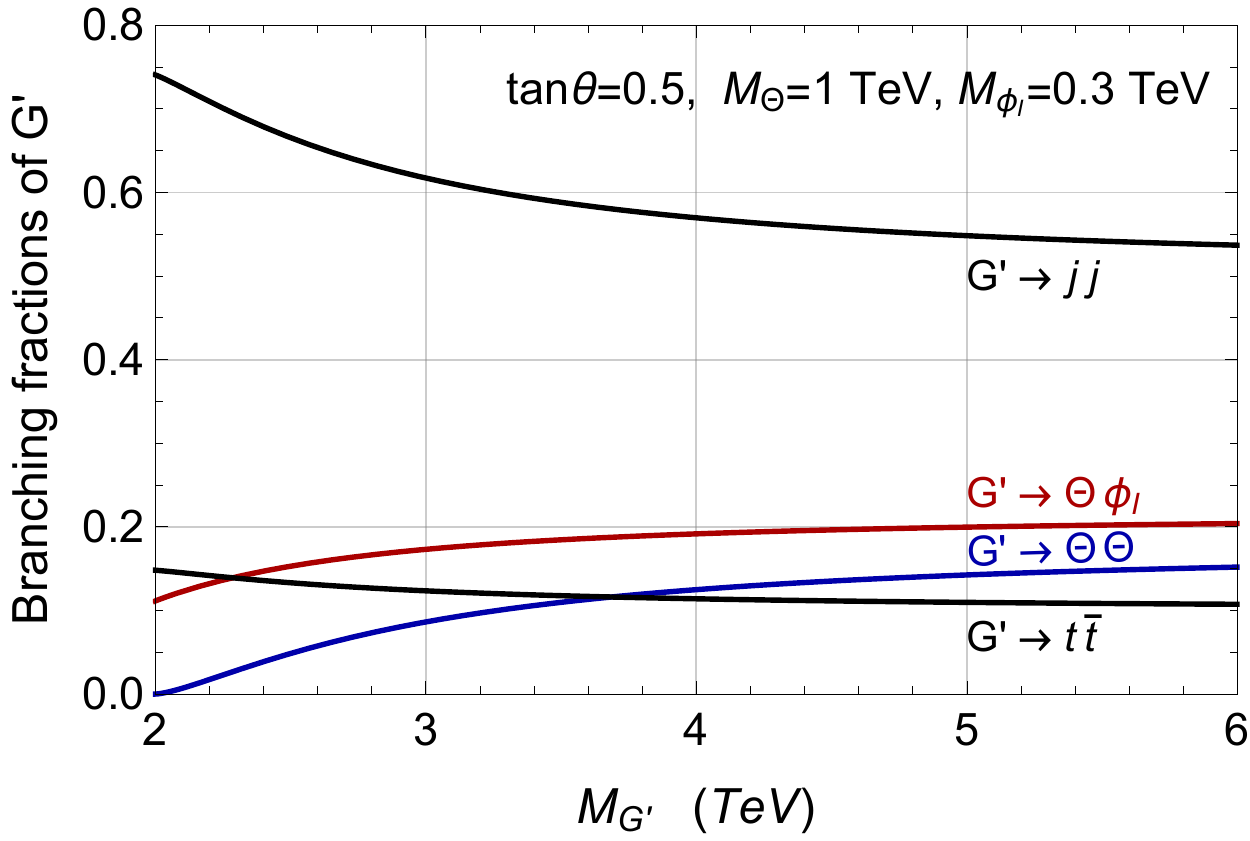}
\caption{Branching fractions of the coloron, $G^\prime$, as a function of its mass, for $\tan\theta = 0.2$ (left panel) or $\tan\theta =  0.5$ (right panel). The scalar masses are fixed at 
$M_\Theta  = 1$ TeV and $M_{\phi_I}  = 300$ GeV.
}
\label{fig:Gprime-branching}
\end{center}
\end{figure}

The coloron with a mass above the scalar pair thresholds, $(M_{G'} > M_{\Theta} + M_{\phi_I} \, , 2 M_\Theta)$ has four 2-body decay channels with tree-level widths:  
\bear
&&\Gamma(G^\prime_\mu \rightarrow \Theta \,\phi_I) =
\frac{\alpha_s}{72\,\tan^2\!\theta} \left(  1 + \tan^2\!\theta \right)^2  M_{G^\prime}   \left( 1 - 2 \frac{M_\Theta^2 + M_{\phi_I}^2 }{M_{G^\prime}^2} + \frac{(M_\Theta^2 - M_{\phi_I}^2)^2 }{M_{G^\prime}^4} \right)^{\! 3/2} \,,      
   \nonumber \\ [2mm]
&& \Gamma(G^\prime_\mu \rightarrow \Theta\, \Theta) = \frac{\alpha_s}{ 32\,\tan^2\!\theta} \left(  1 - \tan^2\!\theta \right)^2 M_{G^\prime}  \left( 1 - \frac{4 M_\Theta^2}{M_{G^\prime}^2} \right)^{\! 3/2} \,,   
 \nonumber \\ [2mm]
&& \Gamma(G^\prime_\mu \rightarrow j j )  = \frac{5 \alpha_s}{6} \tan^2{\!\theta}\; M_{G^\prime} ~~,
 \nonumber \\ [2mm]
&& \Gamma(G^\prime_\mu \rightarrow t \bar t  ) = \frac{ \alpha_s}{6} \tan^2{\!\theta}\; M_{G^\prime}  \left( 1 - \frac{4 m_t^2}{M_{G^\prime}^2} \right)^{\! 1/2} \,.
\label{eq:Gwidths}
\eear
The computation of the partial widths for coloron decays into scalars (first two equations above) is outlined in Appendix B. 
The partial width into a pair of quark jets ($jj$) is summed over five flavors. 
In Figure~\ref{fig:Gprime-branching}, we show the branching fractions of $G^\prime$ as a function of $M_{G'}$.  For $\tan{\theta}$ larger than about 0.45 (as in the right-hand panel), the decays of $G^\prime$ are dominated by the dijet channel, while for a smaller value of $\tan{\theta}$ (as in the left-hand panel), the decays into $\Theta \phi_I$ and $\Theta \Theta$ have large branching fractions. 

%

\subsection{Partial widths of the singlet scalar $\phi_I$}
\label{sec:phiWidth}

The main decay channel of the $\phi_I$ scalar, in the case of the $M_{\phi_I} > M_{\Theta}$ mass ordering, is a 3-body one: $\phi_I \rightarrow j j\,\Theta$ via an off-shell $G^\prime$ with a width
\bear
\hspace*{-0.75cm}
\Gamma(\phi_I \rightarrow j j \Theta) \simeq 5\,\Gamma(\phi_I \rightarrow \bar{t}t\,\Theta) \approx 
 \frac{5\,\alpha_s^2\,M^5_{\phi_I}}{72\pi \,M^4_{G^\prime}}   \left(  1 + \tan^2\!\theta \right)^2  
 {\cal F}(M_{\phi_I}^2  / M_{G^\prime}^2 ,\, M_{\Theta}^2 /M_{G^\prime}^2 ) \,  ,
 \label{eq:3bodyPhi}
\eear
where the function ${\cal F}(x,y)$ is defined in Eq.~(\ref{eq:3bodyFunc}). Comparing Eqs.~(\ref{eq:3bodyTheta}) and (\ref{eq:3bodyPhi}), note that
the order of $M_\Theta$ and $M_{\phi_I}$ is 
reversed in the arguments of  ${\cal F}$.

\begin{figure}[t!]
\begin{center}
\includegraphics[width=0.92\textwidth]{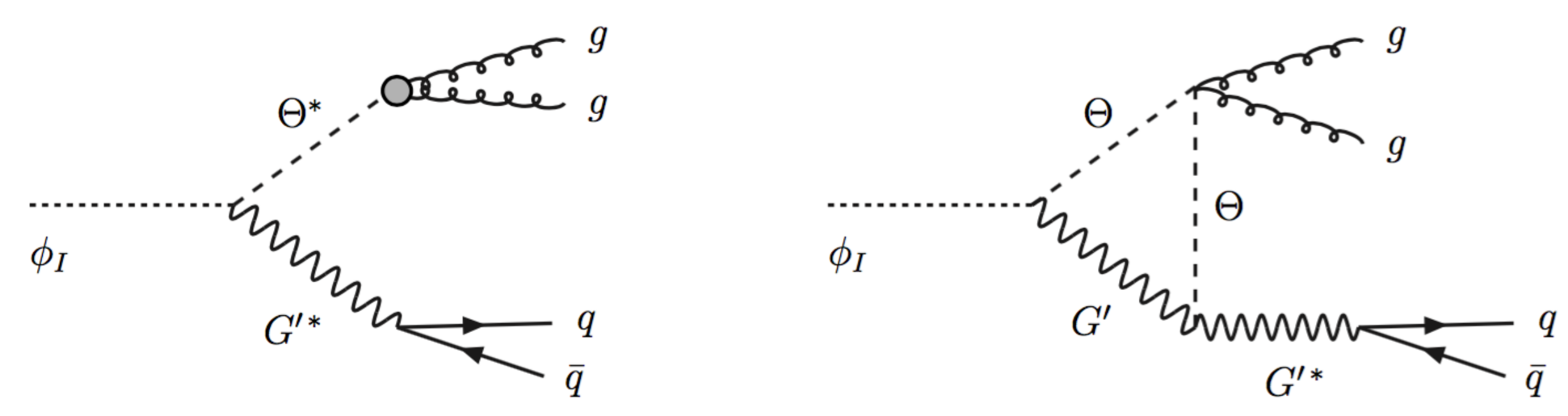} 
\caption{Diagrams responsible for the 4-body decay $\phi_I \to gg q\bar q$  of the $\mathcal{CP}$-odd scalar when $M_{\phi_I} < M_\Theta$. Left diagram: decay through an off-shell coloron and an off-shell $\Theta$
scalar, which couples at one loop to two gluons. Right diagram: 
1-loop decay into a gluon pair and an off-shell coloron (other ways of attaching the gluons to the loop are not shown).
The coloron interactions with scalars are given in Eq.~(\ref{eq:GpScalars}).
}
\label{fig:4body} 
\end{center}
\end{figure}

In the case of the other mass ordering,   $M_{\phi_I} < M_{\Theta}$, 
the leading decay of $\phi_I$ is likely to be a 4-body one into two gluons plus two quarks via the diagrams shown in Figure~\ref{fig:4body}.    
The first diagram there is the decay via an off-shell $G'$ and an off-shell $\Theta$. Note that the $\Theta$ coupling to a gluon pair arises at one loop,
and is described by the dimension-5 operator (\ref{eq:effectiveThetaGluons}), with the coefficient $C_\Theta$ defined in Eq.~\eqref{eq:theta-twogluon}.
 Extracting $f_\Sigma$ from Eq.~(\ref{eq:Gprime-mass}), and using the expressions for $\overline \mu_\Sigma$ and $r_{\mathcal{A}}$ 
given in Eqs.~(\ref{eq:mubar}) and (\ref{eq:r_A}), we find 
\be
C_\Theta \approx  4.61  \left(1 - 0.58 \frac{M_{\phi_I}^2}{M_\Theta^2} +  0.018  \frac{M_\Theta^2}{M_{G'}^2} + \cdots    
\right)   
~~,
\label{eq:CTheta}
\ee
where higher-order terms in  $M_{\phi_I}^2 / M_\Theta^2$ and $M_\Theta^2/M_{G'}^2$ are represented by  the dots.
The effective coupling of $\phi_I$  to gluons and quarks induced by the first diagram in Figure~\ref{fig:4body} is described in the $M_{\phi_I}^2 \ll M_{\Theta}^2, M_{G^\prime}^2$  limit
by a dimension-9 operator:
\be
C_\Theta  \,\frac{g_s^3\,\tan{\theta}}{32\pi^2\,f_\Sigma^2 \,M_{G'}\,M^2_\Theta}\,d^{abc}\, (\partial_\mu \phi_I) \, \bar{q}\gamma^\mu\,T^a q \,  G^b_{\alpha\beta} G^{c\,\alpha\beta} 
~~~ .
\label{eq:phi-decay-operator}
\ee

The second diagram shown in Figure~\ref{fig:4body}, together with four similar 1-loop diagrams where the two gluons are attached in different ways to the $\Theta$ and 
$G'$ internal lines, contribute to the same operator (for $M_{\phi_I}^2 \ll M_{\Theta}^2$), as well as to other dimension-9 operators with different Lorentz contractions.
All these contributions are parametrically of the same order. 
Assuming that $\phi_I$ is much heavier than the QCD scale (so we do not need to take into account hadronization effects),
the 4-body width of $\phi_I$ summed over 5 light quark flavors can be written as
\bear
\Gamma(\phi_I \rightarrow ggq\bar{q}) &=& \frac{2}{189\,(4\pi)^5}\, \left( C_\Theta + C'_\Theta\right)^2  \, \left( \frac{g_s^3\,\tan{\theta}}{32\pi^2\,f_\Sigma^2 \,M_{G'}\,M^2_\Theta}\right)^{\! 2} \, M_{\phi_I}^{11} \,, 
\nonumber \\ [2mm]
&=&  \frac{\,\alpha_s^5}{168\, (12\pi)^4} \, \left( C_\Theta + C'_\Theta\right)^2  \,   \frac{(1 + \tan^2{\theta})^4}{\tan^2{\theta}} \, 
\frac{M_{\phi_I}^{11}}{M_\Theta^4\, M_{G'}^6} \,,
\label{eq:phi-fourbody}
\eear
where $C'_\Theta$ is a function of $M_\Theta/M_{G'}$ that takes into account the 1-loop diagrams of the type shown on the right-hand side of Figure~\ref{fig:4body}.
We expect that $C'_\Theta$ is typically of order one or smaller.
There is also a $\phi_I \rightarrow ggt\bar{t}$ decay, whose width is smaller by a factor of 5 (or more when $M_{\phi_I}$ is not much larger than $2 m_t$).
For $M_{\phi_I}=700$~GeV, $M_{\Theta} = 1$~TeV, $M_{G^\prime}=3$~TeV and $\tan{\theta}=0.3$, the width of  $\phi_I$ is of the order of $5 \times 10^{-13}$~GeV, corresponding to 
a lifetime $\tau^0_{\phi_I} \approx 10^{-12}$~s. 
Note that the lifetime is proportional to $M_{\phi_I}^{-11}$. Thus, a lighter $\phi_I$ produced at the LHC can easily have a long displaced vertex if the 4-body decay is the dominant channel.  
In Section \ref{sec:octet-scalar-2}
we will  estimate the $\phi_I$ decay length in the lab frame.

One may wonder about 2-body decay channels like $\phi_I \rightarrow gg$, as in the case of $\Theta$ discussed earlier (see Section \ref{sec:ThetaWidth}). 
As pointed out below Eq.~\eqref{eq:Sigma-parametrization}, the $\phi_I$ is a $\mathcal{C}$-odd and $\mathcal{P}$-even scalar. Its decay into two gluons is therefore highly suppressed
by charge-conjugation symmetry, requiring electroweak interactions, which enter only at four loops. That symmetry will not forbid $\phi_I$ decays into three gluons. The lowest-dimension operator that is $\mathcal{C}$-conserving appears to be  $\phi_I \, d^{abc}\,G^{a\,\mu}_\nu G^{b\,\alpha}_\mu G^{c\,\nu}_\alpha$, but this operator vanishes because the gluon field strength is antisymmetric in Lorentz indices. At the same dimension, there exist two more operators, $\phi_I \, f^{abc}\,G^{a\,\mu}_\nu G^{b\,\alpha}_\mu G^{c\,\nu}_\alpha$ and $\phi_I \, f^{abc}\,G^{a\,\mu}_\nu G^{b\,\alpha}_\mu \widetilde{G}^{c\,\nu}_\alpha$, both of which are $\mathcal{C}$-violating and thus highly suppressed, again being generated only at four loops.\footnote{The $\mathcal{C}$- and $\mathcal{CP}$-conserving dimension-6 gluon operators without the $\mathcal{C}$-odd $\phi_I$ field were studied in Refs.~\cite{Simmons:1989zs,Dixon:1993xd}.} The next operators are at dimension 9, including $d^{abc}\, (\partial_\mu \phi_I) \, (D_\nu\,G^{a\,\mu\nu}) G^b_{\alpha\beta} G^{c\,\alpha\beta}$, and have the same structure as the one in Eq.~\eqref{eq:phi-decay-operator} after using the field equation for the gluon.

\begin{figure}[t]
\begin{center}
\includegraphics[width=0.9\textwidth]{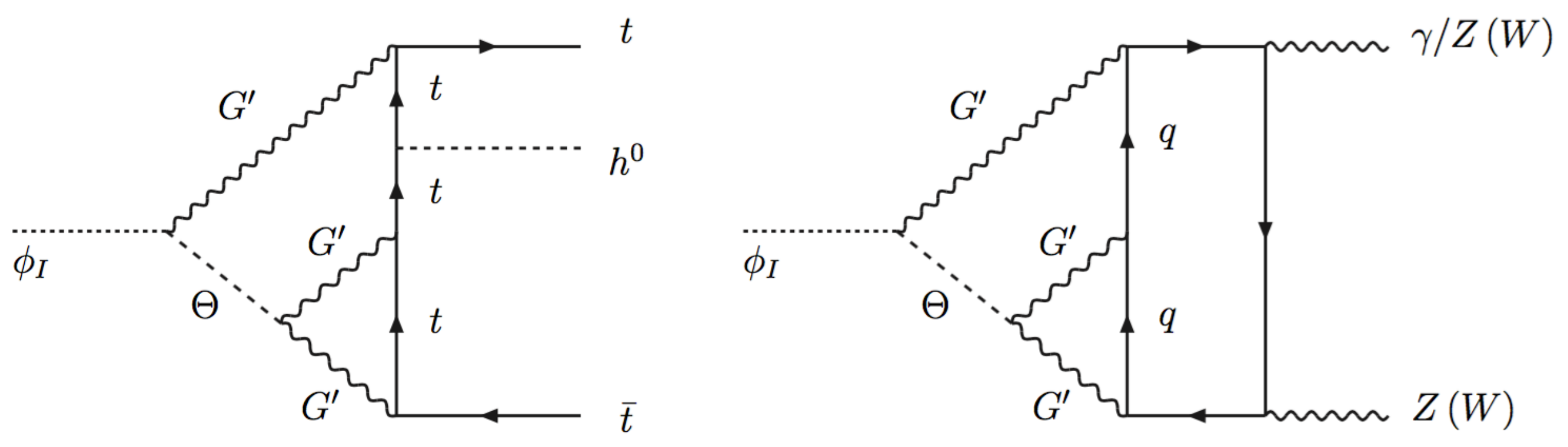} 
\caption{Additional  decays of the $\mathcal{CP}$-odd scalar when $M_{\phi_I} < M_\Theta$. Left diagram: 
$\phi_I \to t \bar t h^0$ at two loops. Right diagram:  $\phi_I \to \gamma Z\,, ZZ\,, W^+ W^-$  at three loops.
}
\label{fig:2loops} 
\end{center}
\end{figure}

Another possible decay of $\phi_I$ is into a quark-antiquark pair, with or without one more boson. To conserve $\mathcal{C}$, the effective operator below electroweak symmetry breaking is $(\partial_\mu \phi_I )\bar{t}\gamma^\mu t$, which is zero after using the fermion equation of motion. The next non-vanishing operator is at dimension 7, $(\partial_\mu \phi_I ) \, HH^\dagger\,\bar{t}\gamma^\mu t$, where $H$ is the Higgs doublet. 
This includes the non-vanishing interaction $\phi_I (\partial_\mu h^0) \bar{t}\gamma^\mu t$, where $h^0$ is the SM Higgs boson. This interaction can be generated by a 
2-loop diagram (see the left panel of Figure~\ref{fig:2loops}), with $\Theta$ and $G^\prime$ running in the loop, and can be written as
\bear
 \eta_1(M_\Theta/M_{G^\prime})\,\tan{\theta}\,(1+\tan^2{\theta})^2\,\frac{10\sqrt{2} \, g_s^5\,y_t^2\,v_{\rm EW}}{9(16\pi^2)^2\,M^3_{G^\prime}}\,  \phi_I (\partial_\mu h^0) \bar{t}\gamma^\mu t  \, ~~.
\eear
Here, $v_{\rm EW} \approx 246$~GeV is the Higgs VEV; $y_t$ is the top quark Yukawa coupling; $ \eta_1$ is a dimensionless coefficient which we expect to be of order one, or slightly smaller; 
the factor $10/9$ comes from color contractions.
For $M_{\phi_I} > m_h + 2 m_t \approx 470$~GeV and ignoring the final state phase-space factor, the 3-body decay width is calculated to be
\be
\Gamma( \phi_I \rightarrow h^0 \, t \, \bar{t} )  \simeq 
 \frac{25\,\alpha_s^5\,\eta_1^2  \,  y_t^4}{162\,(4\pi)^6} \,  \tan^2\!{\theta}\,(1+\tan^2\!{\theta})^4  \,    \frac{v^2_{\rm EW}\,M_{\phi_I}^5}{M_{G'}^6}  ~~.
 \label{eq:phi-three-body-h}
\ee
The decays into lighter quarks are further suppressed by small Yukawa couplings and are negligible. Combining \eqref{eq:phi-fourbody} and \eqref{eq:phi-three-body-h}, the ratio of 
the 2-loop 3-body and 1-loop 4-body decays of $\phi_I$ is
\be
\frac{\Gamma( \phi_I \rightarrow h^0 \, t \, \bar{t} ) }{\Gamma(\phi_I \rightarrow ggq\bar{q}) } \approx  2.6\times 10^{-3} \eta_1^2
 \left( \frac{\tan{\theta}}{0.3} \right)^{\! 4} \, \left( \frac{M_\Theta}{1\,\mbox{TeV}} \right)^{\! 4} \, \left( \frac{700\,\mbox{GeV}}{M_{\phi_I}} \right)^{\! 6}  \, .
\ee
Hence, the 2-loop 3-body decay could be relevant only for a small $M_{\phi_I}$, below around 300 GeV, in which case the collider signature involves a displaced vertex.

As both $\mathcal{C}$ and $\mathcal{P}$ are explicitly broken by the electroweak interactions, there are more 
$\mathcal{C}$ and $\mathcal{P}$-violating operators for $\phi_I$ decaying into SM particles. At the 2-loop level, a Feynman diagram similar to the one on the left side of Figure~\ref{fig:2loops},
with $h^0$ replaced by the hypercharge gauge boson $B_\mu$, generates the following representative dimension-7 operator: 
\be
 \eta_2(M_\Theta/M_{G^\prime})\, \tan{\theta}\,(1+\tan^2{\theta})^2\,\frac{5\,g_s^5\,e}{9\,(16\pi^2)^2\, c_W\, M^3_{G^\prime}}
 \,  (\partial^\mu \phi_I ) \, \overline{q}\gamma^\nu \gamma_5\,q \, B_{\mu\nu} \,,
\ee
where $\eta_2$ is a coefficient of order one or smaller, and $c_W$ is the cosine of the weak mixing angle. 
The ensuing 2-loop 3-body decays, involving  a photon or a $Z$ boson, have a width  
\bear
\Gamma( \phi_I \rightarrow \gamma \, q \, \bar{q} ) \simeq  \frac{c_W^2}{s_W^2} \Gamma( \phi_I \rightarrow Z\, q \, \bar{q} )  
\simeq  \frac{ 5 \alpha\,\alpha_s^5\,\eta_2^2 }{648\,(4\pi)^5}  \, \tan^2\!{\theta}\,(1+\tan^2\!{\theta})^4\;
\frac{M_{\phi_I}^7}{M_{G'}^6}  \, ~,
\label{eq:phi-three-body}
\eear
where we ignored the final state particle masses, and summed over 5 quark flavors.
The ratio of these 2-loop 3-body and  the 1-loop 4-body widths of $\phi_I$ is
\be
\frac{\Gamma( \phi_I \rightarrow \gamma/Z\, q \, \bar{q} ) }{\Gamma(\phi_I \rightarrow ggq\bar{q}) } \approx 1.3 \times 10^{-4}  \eta_2^2 \left( \frac{\tan{\theta}}{0.3} \right)^{\! 4} \, \left( \frac{M_\Theta}{1\,\mbox{TeV}} \right)^{\! 4} \, \left( \frac{700\,\mbox{GeV}}{M_{\phi_I}} \right)^{\! 4}  \, .
\ee
There are also decay channels involving the $W$ boson, $\phi_I \rightarrow W^\pm\, q \, \bar{q}'$ with comparable branching fractions. Thus, the 2-loop 3-body widths with an electroweak boson in the final state 
are probably  too small to be relevant for collider studies. 

At the 3-loop level, $\phi_I$ couples to two electroweak bosons or two SM fermions. A diagram of this type is shown on the right-hand side of 
Figure~\ref{fig:2loops}.
The $\mathcal{CP}$-conserving operators that couple the $\mathcal{CP}$-odd scalar to two electroweak gauge bosons include 
\bear
\hspace*{-3mm}
 \eta_3(M_\Theta/M_{G^\prime})\, \tan{\theta}\,(1+\tan^2{\theta})^2\,
 \frac{ 5\,g_s^5\,e^2}{ 2 (16\pi^2)^3 \, M_{G^\prime}} \, \phi_I\,\left( \frac{1}{s_W^2}\,W^{i}_{\mu\nu}\,\widetilde{W}^{i\,\mu\nu} -\frac{1}{c_W^2}\,B_{\mu\nu}\,\widetilde{B}^{\mu\nu} \right)  ,
 \label{eq:3loopOps}
\eear
where $\eta_3$ again is a coefficient of order one or smaller, which depends on the $\phi_I$, $\Theta$ and $G'$ masses, and can be found by computing the 3-loops diagrams.
The coefficient of the above operator for the hypercharge gauge boson includes a factor of $-3/2$, which accounts for the sum over the squared hypercharges of the six quark flavors running in the loop. The coefficient for the $W^i$ gauge bosons comes from three generations of quarks and $1/2$ for the normalization of $SU(2)_W$ generators. The overall coefficient in (\ref{eq:3loopOps}) includes 
a color factor of 10/3.
Interestingly, after rotating the basis to $\gamma$, $Z$ and $W^\pm$,  there is no 3-loop coupling of $\phi_I$ to two photons. For other combinations of two gauge bosons, 
the 3-loop 2-body decays have widths
\bear
 \Gamma(\phi_I \! \rightarrow \! \gamma Z)  \! \! & \! \simeq \! &  \! \! \frac{s_W^2}{ c_W^2}  \Gamma(\phi_I \! \rightarrow \! W^+W^-)   \simeq   \frac{2 s_W^2 c_W^2}{1 - 4 s_W^2 c_W^2 }   \Gamma(\phi_I \! \rightarrow \! ZZ)  
\nonumber \\ [2mm]
& \simeq  &  \!
 \frac{25 \, \alpha^2  \alpha_s^5 \, \eta_3^2 }{ 2(4\pi)^6 \,  s_W^2 c_W^2 }
\tan^2\!{\theta}  \, (1+\tan^2{\theta})^4   \,
 \frac{M_{\phi_I}^3}{M_{G'}^2}  \,  ~~.
 \label{eq:decay-two-boson}
\eear
The ratios of these 3-loop 2-body widths to the 1-loop 4-body width of $\phi_I$ given in Eq.~(\ref{eq:phi-fourbody}) are
\be
\frac{ \! \Gamma( \phi_I \rightarrow  \gamma Z, WW ,ZZ) \! }{\Gamma(\phi_I \rightarrow ggq\bar{q}) } \approx (0.20,0.66, 0.16) \eta_3^2 \left( \frac{\tan{\theta}}{0.3} \right)^{\! 4} \left( \frac{M_{G'}}{3\,\mbox{TeV}} \right)^{\! 4} \left( \frac{M_\Theta}{1\,\mbox{TeV}} \right)^{\! 4} \left( \frac{700\,\mbox{GeV}}{M_{\phi_I}} \right)^{\! 8}  .
\label{eq:two-body-over-four-body}
\ee
If $\eta_3 = O(1)$,  the 2-body decays into two electroweak bosons become the leading decay channels of $\phi_I$
 for $M_{\phi_I}$ below around 700 GeV, or for $M_{G'} M_\Theta $ above 3 TeV$^2$. 
If  $\eta_3$ turns out to be much smaller, $\eta_3^2 = O(10^{-3})$, then the  2-body decays become dominant for $M_{\phi_I} \lesssim 300$ GeV when $M_{G'} M_\Theta \sim  3$ TeV$^2$. 

At the 3-loop level ({\it e.g.}, the left diagram of Figure~\ref{fig:2loops} with an additional $W$ loop attached to the top quark lines and no $h^0$), 
the $\mathcal{CP}$-conserving couplings to two fermions contain the dimension-5 operator
\be
 \eta_4(M_\Theta/M_{G^\prime})\, \tan{\theta}\,(1+\tan^2{\theta})^2\,\frac{5\,g_s^5\,e^2}{9 (16\pi^2)^3\, s_W^2 \, M_{G^\prime}} (\partial_\mu\phi_I) \, \overline{t}\,\gamma^\mu\gamma_5 t \,.
\ee
After using integration by parts and fermion equation of motion, it is easy to show that the coupling is proportional to the fermion mass, so the top quark (if kinematically allowed) is the most important channel. Ignoring the final-state fermion masses, this 3-loop 2-body decay width is then
\bear
\Gamma(\phi_I \rightarrow t \overline{t} ) \simeq 
\frac{25\,\alpha^2\,\alpha_s^5\,\eta_4^2}{ 81(4\pi)^6 \, s_W^4}  \tan^2\!{\theta}  \, (1+\tan^2{\theta})^4 \,    \frac{m_t^2\,M_{\phi_I}}{M_{G'}^2} \,.
\label{eq:decay-two-top}
\eear
The ratio of this decay mode over the 1-loop 4-body one is
\be
\frac{\Gamma( \phi_I \rightarrow t \bar{t} ) }{\Gamma(\phi_I \rightarrow ggq\bar{q}) } \approx 1.0\times 10^{-3} \eta_4^2 \left( \frac{\tan{\theta}}{0.3} \right)^{\! 4}\, \left( \frac{M_{G'}}{3\,\mbox{TeV}} \right)^4\, \left( \frac{M_\Theta}{1\,\mbox{TeV}} \right)^{\! 4} \, \left( \frac{700\,\mbox{GeV}}{M_{\phi_I}} \right)^{\! 10}  \,.
\ee
Comparing the two 3-loop widths in \eqref{eq:decay-two-boson} and \eqref{eq:decay-two-top}, the fermion pair channel is suppressed by $m_t^2/M^2_{\phi_I}$ for comparable $\eta_3$ and $\eta_4$. 

In conclusion, for $M_{\phi_I}$ not dramatically below $M_{G'}$ and $M_\Theta$, the main decay of $\phi_I$ is the 1-loop 4-body channel $\phi_I \rightarrow ggq\bar{q}$ in \eqref{eq:phi-fourbody}, otherwise the main channels appear to be $\phi_I \rightarrow \gamma Z, W^+ W^-, ZZ$ with the widths estimated in \eqref{eq:decay-two-boson}. 

\bigskip\medskip

\section{Production of scalars at the LHC}
\label{sec:production} \setcounter{equation}{0}

The ReCoM is a relatively simple gauge theory. It contains only four new particles beyond the SM: the spin-1 coloron, the color-octet scalar $\Theta$, and two singlet scalars.
Nevertheless, there are several processes at the LHC that lead to the production of these particles.

\begin{figure}[t]
\begin{center}
\includegraphics[width=0.65\textwidth]{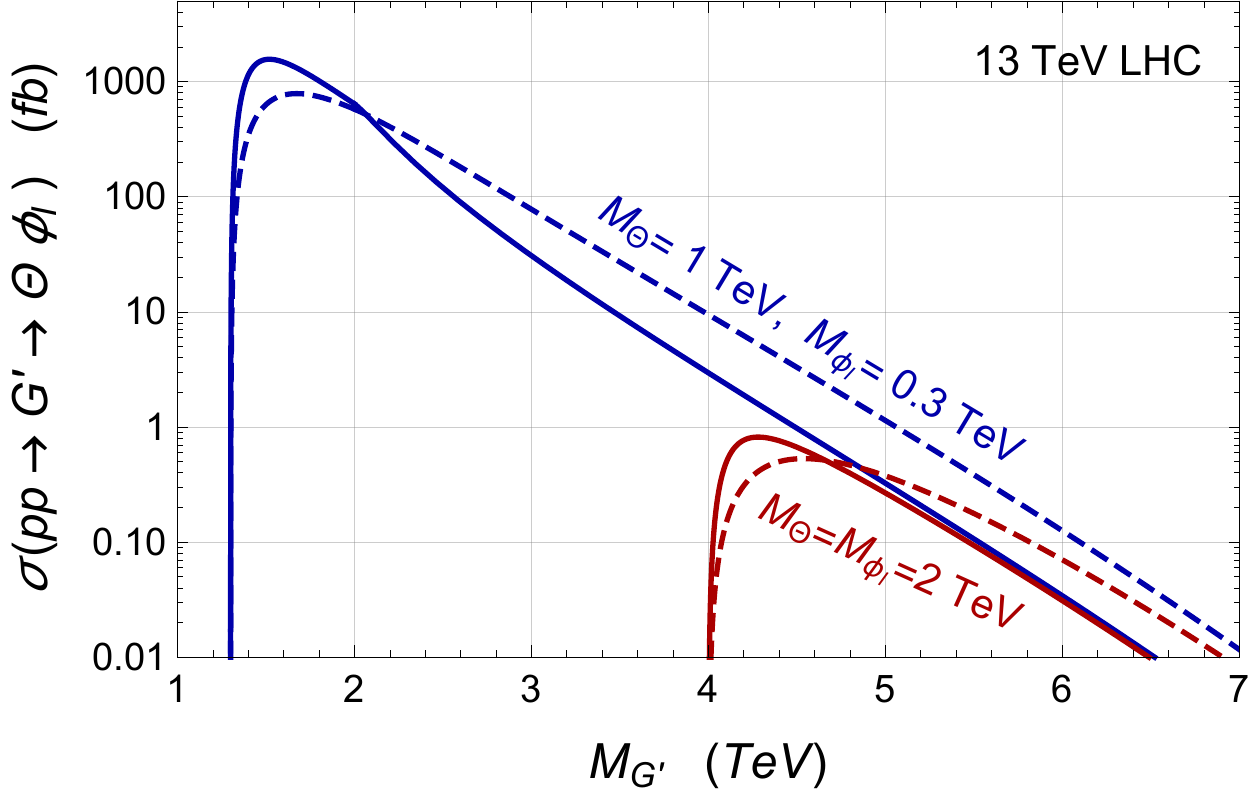} 
\caption{Cross section for the $pp  \to G' \to \Theta \, \phi_I$ process at $\sqrt{s}= 13$ TeV as a function of the coloron mass for $\tan\theta = 0.2$ (solid lines) or 0.5 (dashed lines).
The scalar masses are fixed at $M_\Theta = 1$ TeV and $M_{\phi_I} = 300$ GeV (blue lines), or at  $M_\Theta = M_{\phi_I} = 2$ TeV (red lines). }
\label{fig:prodthetaphi}
\end{center}
\end{figure}

A peculiar production process is $q\bar q \to G' \to \Theta \, \phi_I$, which is resonant production of two scalar particles of different masses and carrying different color charges.
In the narrow width approximation, the tree-level $G^\prime$ parton-level production cross section is given by  
\be
\sigma(q\bar{q} \rightarrow G^\prime) \approx \frac{8\pi^2 \alpha_s \, \tan^2\! {\theta}}{9 M_{G^\prime}}    \; \delta(\sqrt{\hat{s}} - M_{G^\prime}) ~~. 
\ee
Next-to-leading order corrections to coloron production have been computed in \cite{Chivukula:2011ng,Chivukula:2013xla}, and can be approximated by a multiplicative $K$-factor 
of about 1.2.
At the 13 TeV LHC, the $G'$ production cross section is approximately $(20, 1.8, 0.20)$~pb for $M_{G^\prime}= 2, 3, 4$~TeV. 
Using the MSTW parton distribution functions~\cite{Martin:2009iq}, $K = 1.2$, and the branching fraction for $G' \to \Theta \, \phi_I$ (see Figure~\ref{fig:Gprime-branching}), we find 
the cross section for  $\Theta \, \phi_I$ production shown in Figure~\ref{fig:prodthetaphi}. 

The color-octet scalar $\Theta$ can be pair-produced at hadron colliders through two very different processes. The first one, which we loosely refer to as QCD production, is due to the 
 $\Theta$ coupling to the gluons, and is governed by the QCD gauge coupling $g_s = \sqrt{ 4 \pi \alpha_s} \approx 1$ (here $\alpha_s$ is the strong coupling constant at a scale of order $2 M_\Theta$).
There are four types of Feynman diagrams that contribute to QCD production at leading order \cite{Bai:2010dj}: three diagrams from the $gg$ initial state (via an $s$-channel gluon, a $t$-channel $\Theta$, 
or a $gg\Theta\Theta$ coupling) and one diagram from $q\bar q$ initial states  (via an $s$-channel gluon). 
At the LHC, the gluon-initiated processes have a  larger rate than the  $q\bar q$ ones, unless $\Theta$ is very heavy.

\begin{figure}[t]
\begin{center}
\includegraphics[width=0.65\textwidth]{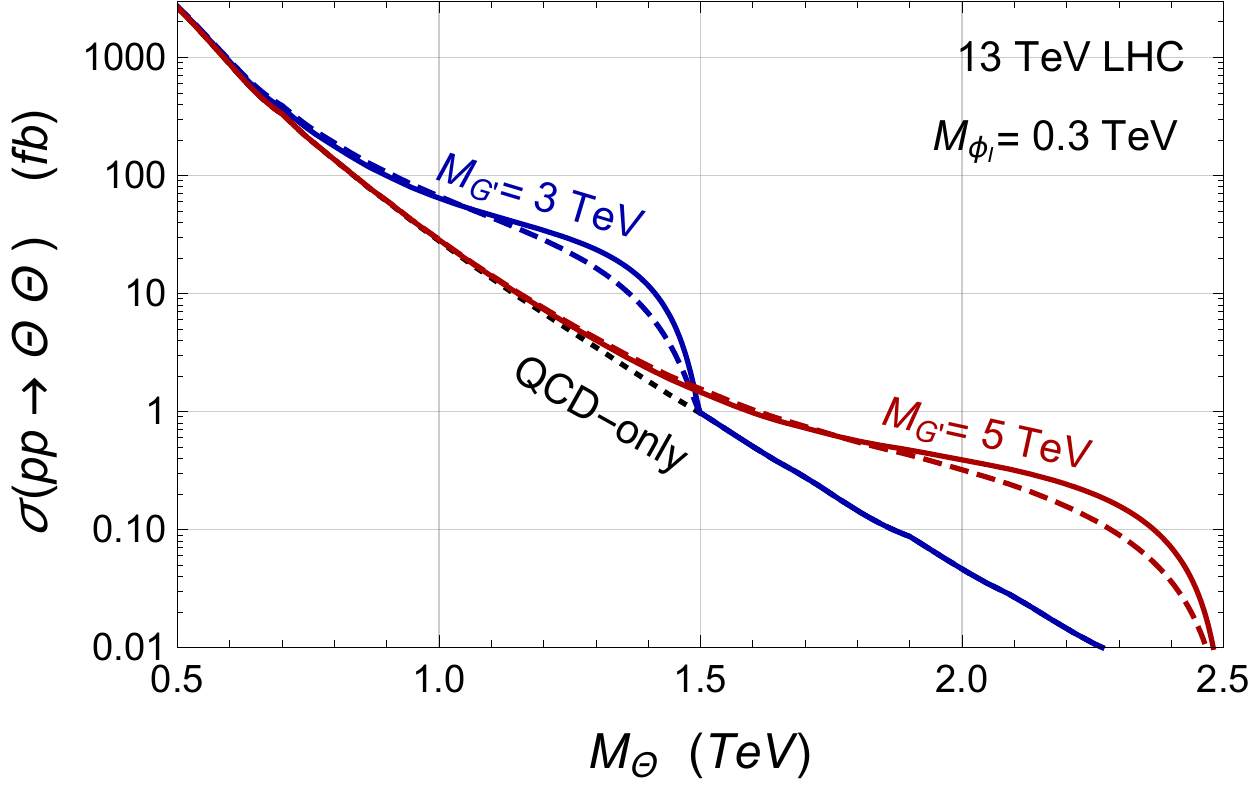} 
\caption{Leading-order cross section for $ \Theta \Theta$ production (due to an $s$-channel $G'$, QCD and interference) at the 13 TeV LHC as a function of the $\Theta$ mass. 
The coloron mass is fixed at $M_{G'} = 3$ TeV (blue lines) or 5 TeV (red lines), while $\tan\theta = 0.2$ (solid lines) or 0.5 (dashed lines). The dependence on the neutral-scalar 
mass (fixed here at $M_{\phi_I} = 300$ GeV) is mild.
The dotted line represents the QCD contribution.}  
\label{fig:prodThetaTheta}
\end{center}
\end{figure}

The second process that leads to $\Theta$ pair production involves an $s$-channel coloron. At leading order this is due to quark-antiquark initial states, so it interferes only with the $q\bar q$-initiated QCD production (this represents between 10\% and 30\% of the nonresonant $\Theta\Theta$ production rate for $M_\Theta$ in the 1--2.5 TeV range). The interference effects are thus small, are further suppressed when the coloron is a narrow resonance, and may be relevant only for a small range of parameters where the amplitudes for QCD and resonant productions are comparable. The total  $\Theta\Theta$  production cross section at the 13 TeV LHC is shown in Figure~\ref{fig:prodThetaTheta} as a function of the $\Theta$ mass, for $M_{G'}= 3$ TeV or 5 TeV, $\tan\theta = 0.2$
or 0.5, and $M_{\phi_I} = 300$ GeV. The dependence on $M_{\phi_I} $ (due to the $G' \to \Theta \, \phi_I$ branching fraction) is negligible when this parameter is much smaller than $M_\Theta$.

\begin{figure}[t]
\begin{center}
\includegraphics[width=0.7\textwidth]{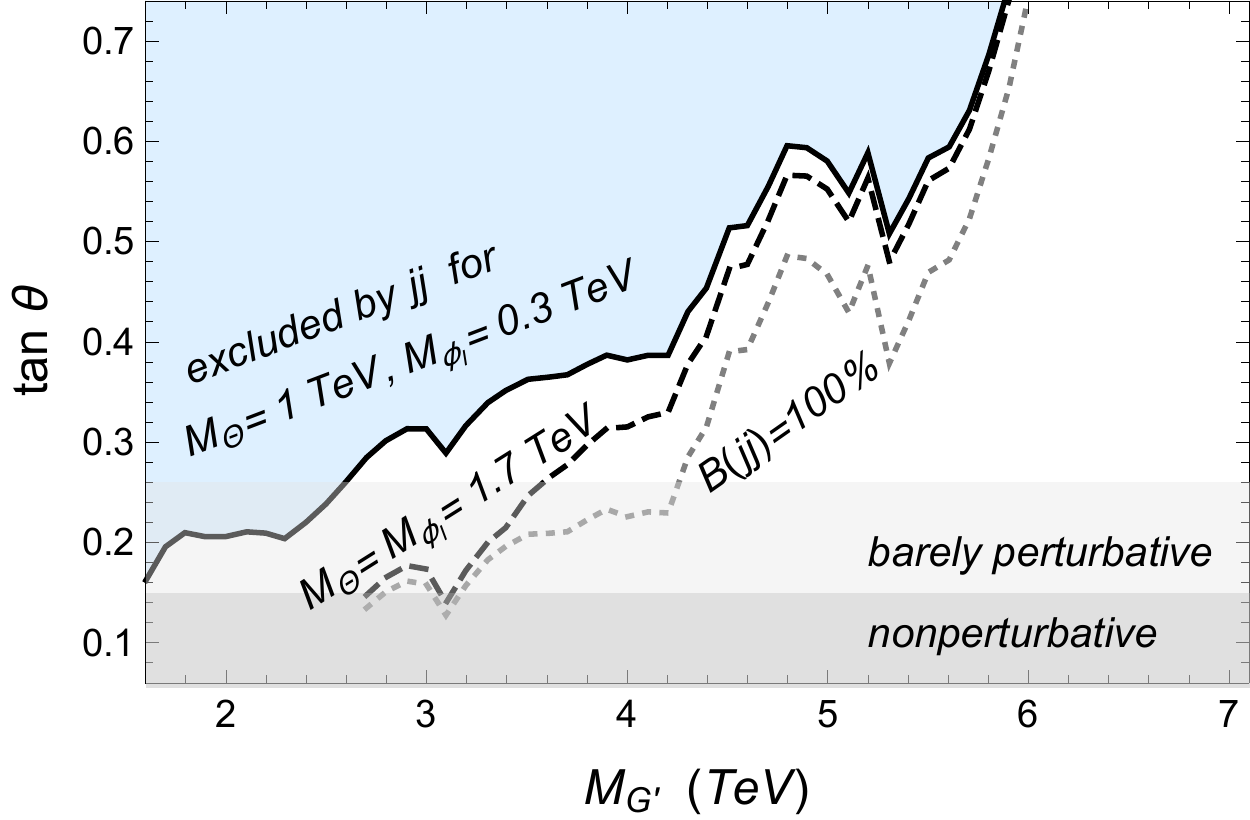} 
\caption{Excluded regions of the coloron parameter space. Region above solid black line is ruled out, for $M_\Theta = 1$ TeV and $M_{\phi_I}= 300$ GeV,
by $jj$ resonance searches, assuming that  $pp \to G' \to jj$ is the only new-physics process contributing to the dijet final state. Similarly, the region above 
the dashed line is ruled out for $M_\Theta = M_{\phi_I} = 1.7$ TeV. The region above the dotted line is ruled out when the effective $jj$ branching fraction is 
100\% ({\it e.g.},  scalar masses are above $M_{G'}/2$ and the boosted $t\bar t$ signal mimics a $jj$ one).
In the region labelled ``barely perturbative", the expansion parameter for loops that involve 
the coloron self-coupling is relatively large [$3h_2^2/(16\pi^2)$ between 1/3 and 1] so that the tree-level result is expected to get sizable corrections.
}
\label{fig:theta-exclusion}
\end{center}
\end{figure}

There are also two classic  processes, 
 $q\bar q \to G' \to jj$ and $q\bar q \to G' \to t \bar t$, which have been searched for by the ATLAS and CMS experiments, and can be used to exclude a certain region
 of the parameter space. 
After taking into account the acceptance of around 0.6~\cite{CMS:2017xrr}, the CMS collaboration has set an upper constraint on the cross section times branching fraction 
($\sigma  B$) from dijet resonance searches with 
36 fb$^{-1}$ of data (similar constraints are also imposed by the ATLAS collaboration~\cite{Aaboud:2017yvp}). For $M_{G^\prime}=(2, 3, 4)$~TeV, the constraint is $\sigma(pp \rightarrow G^\prime) B(G^\prime \rightarrow jj) \leq (100, 20, 10)$
fb~\cite{CMS:2017xrr}. For a small mixing of $\tan{\theta} = 0.2$, $B(G^\prime \rightarrow jj)$ could be as small as 5\% such that the constraint on the coloron mass is around $M_{G^\prime} \gtrsim 2$~TeV.

 In Figure~\ref{fig:theta-exclusion}, we show the constraints on the mixing parameter $\tan{\theta}$ for different coloron masses from the dijet resonance searches. We also note that the $t\bar{t}$ resonance searches are less sensitive because of the smaller branching fraction\cite{Sirunyan:2017uhk}. It is clear from the left panel of Figure~\ref{fig:Gprime-branching} that the main decay channels of the coloron for small $\tan\theta$ are into the scalars in the ReCoM, which provides novel signatures at the LHC and will be the focus of next section.  

The second singlet scalar, $\phi_R$, cannot be produced in coloron decays, as it only couples to $G'_\mu G^{\prime \mu}$. This coupling allows the 
non-resonant production of $\phi_R$ in association with a coloron. Single production of $\phi_R$ also proceeds through a gluon fusion process induced 
at one loop, with separate contributions from the coloron and the $\Theta$ scalar running in the loop.
If $\phi_R$ is heavy enough, it can decay into a pair of $\phi_I$ or $\Theta$ scalars
due to the trilinear couplings in Eq.~(\ref{eq:3scalar}). The  $\phi_R$ scalar may also decay into pairs of heavy SM particles due to its mixing with the SM Higgs boson,
as studied in \cite{Chivukula:2013xka, Chivukula:2014rka}.

Another class of processes is based on coloron pair production through its gluon couplings. Even though this production is non-resonant and 
kinematically suppressed by the presence of two heavy colorons, its cross section relies solely on the QCD coupling and 
is above 1 fb for a coloron mass below about 2.5 TeV \cite{Dobrescu:2007yp}. 
Depending on the decay channels, this may be large enough for the high-luminosity run of the LHC.
The coloron decays into scalar pairs in this case leads to intermediate states involving $4\Theta$ or $3\Theta + \phi_I$ or $\Theta\phi_I\Theta\phi_I$,
which give rise to complicated final states with high jet multiplicity.

In the next section we will focus on the final states that arise from $\Theta \phi_I$  or  $\Theta\Theta$  production, which have larger cross sections. 

\bigskip

\section{Novel LHC signatures}
\label{sec:LHC} \setcounter{equation}{0}

The ReCoM predicts rich collider phenomena with many complicated final states involving 
hadronic activity and other objects.
At the LHC, the single production of $\Theta$ or $\phi_I$ is suppressed due to loop factors. Therefore, we will concentrate on the pair-productions $pp \rightarrow \Theta \Theta$ and $pp \rightarrow \Theta \phi_I$ and point out the novel signatures in the ReCoM. 

To study the current experimental bounds and interesting signatures that have not been searched for at the LHC, we consider two types of mass orderings: $M_{G^\prime} > M_{\phi_I} > M_\Theta$ and  $M_{G^\prime}> M_\Theta >  M_{\phi_I}$. In the following, we will study the leading collider signatures for each case and point out the discovery potential.

\subsection{Light color-octet scalar: $M_{G^\prime} > M_{\phi_I} > M_\Theta$}
\label{sec:octet-scalar}

For this parameter region, the color-octet scalar $\Theta$ mainly decays into two gluons ($\Theta \rightarrow gg$), while the singlet 
scalar predominantly decays via an off-shell coloron into a  3-body final state: $\phi_I \rightarrow j j \Theta$ with the two jets originating from quarks. 
After the subsequent $\Theta$ decaying to two gluons, $\phi_I$ behaves as a 4-jet resonance with a 2-body sub-resonance.

\begin{figure}[b!]
\begin{center}
\vspace*{0.25cm}
\includegraphics[width=0.75\textwidth]{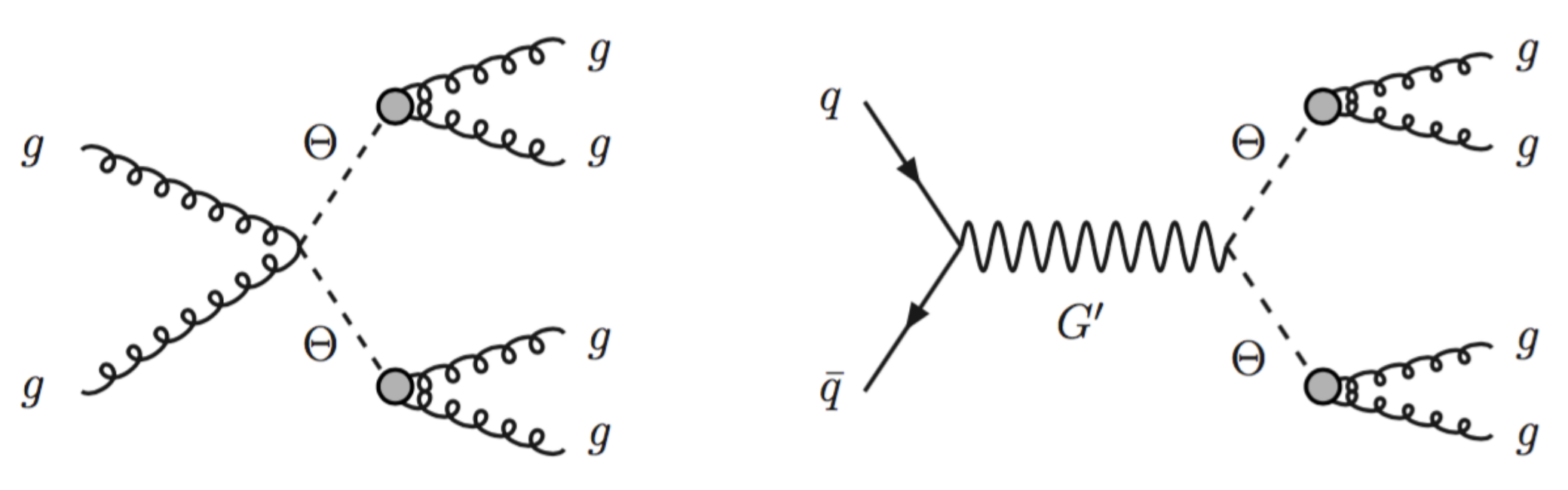} 
\caption{Pair production of the color-octet scalar $\Theta$, through QCD (left-hand diagram; related diagrams with only trilinear vertices are not shown), and coloron decay (right-hand diagram), followed by $\Theta$  decays into two gluons at one loop. The branching fraction for $\Theta \to gg$ is close to 100\% when $M_\Theta < M_{\phi_I}$. }
\label{fig:ThetaPair}
\end{center}
\end{figure}

Starting from the pair production of $pp \rightarrow \Theta \Theta \rightarrow 4g$, the final state contains four jets with a pair of dijet resonances of equal mass, as can be seen in Figure~\ref{fig:ThetaPair}. This final state with a pair of dijet resonances has been studied in Refs.~\cite{Chivukula:1991zk,Dobrescu:2007yp,Kilic:2008pm,Bai:2010dj} and searched for at the LHC by both CMS~\cite{Chatrchyan:2013izb,Khachatryan:2014lpa} and ATLAS~\cite{ATLAS:2016sfd,ATLAS:2017gsy}. Using the latest result from ATLAS with 36.7 fb$^{-1}$ data~\cite{ATLAS:2017gsy},  the upper limit on the cross section for production  of a color-octet scalar pair
is around 0.4~pb for $M_\Theta =700$~GeV, which approximately matches the production cross section predicted in the ReCoM, shown 
in Figure~\ref{fig:prodThetaTheta}. So, for this region of parameter space, the limit on the color-octet scalar mass is $M_\Theta \gtrsim 700$~GeV.

As shown in Figure~\ref{fig:prodThetaTheta}, the $\Theta$ pair production is typically 
dominated by the coloron-mediated process when $M_{G'} > 2 M_\Theta$. Thus, besides searching for a pair of dijet resonances, one could also search for a resonance in the invariant mass spectrum of all four jets (for a related collider study with $b$ jets at the Tevatron, see \cite{Bai:2010dj}).

For a hierarchical mass spectrum with $M_{G^\prime} \gg M_{\Theta}$, the $\Theta$ particle from $G^\prime$ decays could be highly boosted and behave as a jet with 2-prong substructure at the LHC. 
The latter has an invariant mass given by $M_\Theta$, and we label it by $J_\Theta$.
As a rough estimate, the angular separation of the two gluon-jets from $\Theta$ decays has a lower bound approximately given by \cite{Butterworth:2008iy}
\be
\Delta R_{gg} \simeq \frac{2\,  M_{\Theta}}{p_T(\Theta)} \simeq 4\, \frac{M_{\Theta}}{M_{G^\prime}}  ~~.
\label{eq:DeltaR}
\ee
For instance, one could enlarge the jet-finding-algorithm radius to $R \sim 0.8$ for $M_{\Theta}=1$~TeV and $M_{G^\prime}=5$~TeV and use additional jet-substructure techniques to search for the signal in this case~\cite{Bai:2011mr}.  

If the scalars are even more boosted such that the two gluon-jets are within the jet cone used in regular 
dijet searches, then the $pp\to G' \to \Theta\Theta \to J_\Theta J_\Theta$ process will effectively appear as a dijet resonance. The effective dijet branching fraction of the coloron would then 
be larger, and  the constraint from dijet resonance searches would be stronger (see Figure \ref{fig:theta-exclusion}).
However, using a typical $\Delta R_{gg} = 0.4$ in  Eq.~(\ref{eq:DeltaR}) would require $M_\Theta/M_{G'} \lesssim O(0.1)$. This in conjunction with the 
lower limit on $M_\Theta$ would push $M_{G'}$ above 7 TeV, where the production cross section becomes too small.

\begin{figure}[t]
\begin{center}
\includegraphics[width=0.45\textwidth]{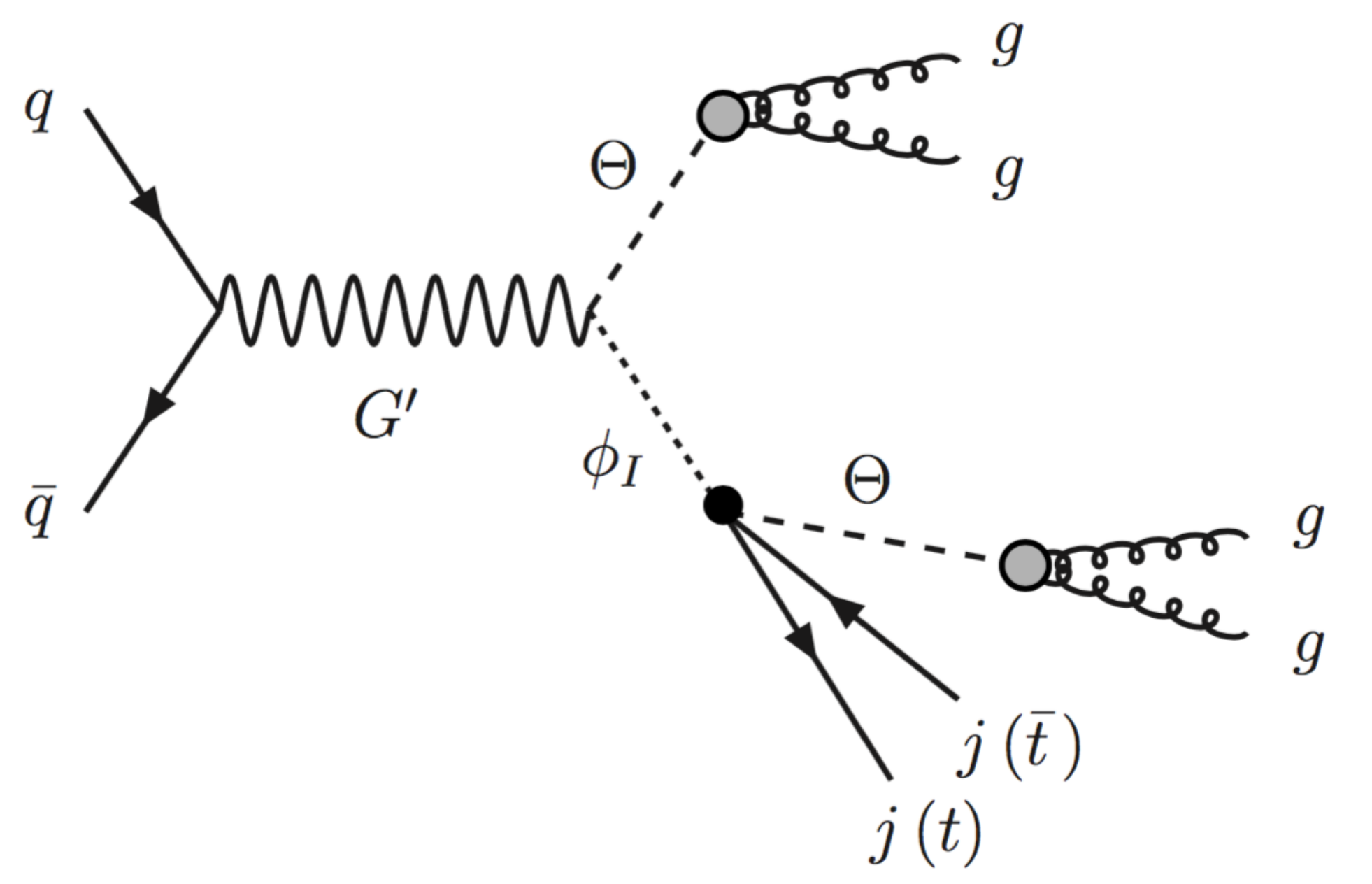} 
\caption{Resonant production of a coloron  that decays into an octet scalar $\Theta$ and a singlet scalar $\phi_I$, which then decays (for $M_\Theta < M_{\phi_I}$) via an off-shell coloron into a $\Theta$ plus either two quark jets or $t\bar t$.}
\label{fig:ThetaPhi}
\end{center}
\end{figure}

For the singlet scalar, $\phi_I$, the main production is together with a $\Theta$ scalar via $G^\prime$ decays. The corresponding Feynman diagram for its production and decay is shown in Figure~\ref{fig:ThetaPhi}. When it is produced at the LHC, the $\phi_I$ undergoes a 3-body decay into $\Theta$ plus a quark pair via an off-shell coloron. When the quark is not the top, the quark pair hadronizes into two jets.
After the two $\Theta$'s decay, there are altogether six jets in the final state, with two of them having an invariant mass of $M_\Theta$ and four of them an invariant mass of $M_{\phi_I}$:
$ pp \to G' \to \Theta\phi_I \to \Theta (\Theta q \bar q) \to (gg) \, ((gg) q \bar q) $. 
The total invariant mass of all six jets should match the coloron mass $M_{G^\prime}$. To our knowledge, there are no dedicated experimental searches to cover this signature. A less sensitive but related search is looking for microscopic black holes at the 13 TeV LHC via multiple jets~\cite{CMS-PAS-EXO-15-007} with $2.3$~fb$^{-1}$ luminosity. Based on a parton-level simulation, we conclude that the upper bound on the cross section for six or more  jets does not constrain the ReCoM.  

Searches for the signal of multiple resonances in 6-jet final states would suffer from a combinatorial background. For instance, there are $C^2_6 C^2_4=90$ dijet pairs if
the search is designed to find the two dijet resonances associated with $\Theta$. This combinatorial issue would likely make the reconstructed $\Theta$ resonance very broad.  

Instead of reconstructing the $\Theta$ or $\phi_I$ resonances, a simple strategy is to use the invariant mass of all six jets to search for  the $G^\prime$ resonance. 
Even then, additional jets from initial state radiation would complicate the reconstruction of the resonance. In addition, 
the jet energy resolution  in the presence of a large number of jets would make the resonance broad.

The situation is dramatically different when $M_{G^\prime} \gg  M_{\phi_I} >  M_{\Theta}$. 
The signal arising from the asymmetric decay of a coloron produced in the $s$-channel, $pp\to G' \to \Theta \phi_I $,  is then a ``dijet" resonance, with one of the jets ($J_\Theta$) having a 2-prong substructure, 
and the other jet (of mass $M_{\phi_I}$ and labelled by $J_{jj \Theta}$) having a 4-prong substructure (similar boosted objects are discussed in \cite{Aguilar-Saavedra:2017rzt,Aguilar-Saavedra:2018xpl}). In this case, the effective dijet branching fraction is closer to 100\% (provided that each multi-prong jet fits inside a $\Delta R = 0.4$ cone), so that the constraint from dijet resonance searches approaches the dotted line in Figure \ref{fig:theta-exclusion}. 

The case where $M_{G^\prime} >  M_{\phi_I}  \gg  M_{\Theta}$ would still be phenomenologically different.
Note, however, that this case is not consistent with the upper limit $M_{\phi_I} < 2.1 M_{\Theta}$ derived in Appendix A, which 
is based on the assumption that the color-preserving vacuum is the global minimum of the scalar potential.

Nevertheless, there are interesting intermediate cases, where $M_{G^\prime}$ is considerably 
 larger than $M_{\phi_I} + M_{\Theta}$ such that the primary $\Theta$ is moderately boosted, while the two quark jets and the $\Theta$ arising from the 
$ \phi_I$ decay have only a small boost. 
An example of mass spectrum that leads to the above situation is $M_\Theta \approx 1$ TeV, $M_{\phi_I} \approx 2$ TeV and $M_{G^\prime} \approx 6 $ TeV.
The signal is a $J_\Theta$,   which in this case is a wide jet of  $\Delta R_{gg} \approx 0.8$, 
and four other jets that reconstruct the $\phi_I$ mass. The reconstruction of the resonances would be improved by techniques that distinguish 
between gluon jets and quark jets.

A separate class of signatures arises from the $\phi_I \to t\bar t \, \Theta$ decay (see Figure~\ref{fig:ThetaPhi}). Even though the branching fraction for this 3-body decay is 1/6, or smaller 
when $M_{\phi_I}-M_{\Theta}$ is near $2 m_t$,  the more complicated final states associated with top quarks may be used to reduce the backgrounds.
For example, one may use the leptonic decay of a $W$ from the  $pp\to G' \to  \phi_I \Theta \to   W^+W^- b\bar b+ 4j$ process. 

If $M_{G^\prime} \gg M_{\phi_I} > M_{\Theta}$, then the $\phi_I \to t\bar t \, \Theta \to  t\bar t \, jj $ system is also boosted, giving rise to an object (we label it by  $ J_{t\bar t \Theta}$)
that includes multi-prong jet substructure and sometimes nonisolated leptons and missing energy.
The signal is then $pp \to G' \to   J_\Theta  J_{t\bar t \Theta}  $.

\bigskip

\subsection{Light singlet scalar: $M_{G^\prime} > M_\Theta > M_{\phi_I}$}
\label{sec:octet-scalar-2}

Let us now study the mass ordering where the singlet scalar $\phi_I$  is lighter than the color-octet scalar $\Theta$. In that case, $\Theta$ has 2-body decays into two gluons and 3-body decays into $j j \phi_I$ with the branching fractions depending on $\tan\theta$ and on
the $M_{\Theta}/M_{G^\prime}$ and $M_{\phi_I}/M_{G^\prime}$ mass ratios, as follows from Eqs.~(\ref{eq:decaygg}) and (\ref{eq:3bodyTheta}) (see also 
Figure~\ref{fig:theta-branching}).

The $\phi_I$ undergoes 
4-body 1-loop decays into two gluons and a quark-antiquark pair,  or 2-body 3-loop decays into $W^+W^-$, $\gamma Z$ or $ZZ$.
The latter are dominant for $M_{\phi_I} < M_0$ where the mass scale $M_0$ can be derived from Eq.~(\ref{eq:two-body-over-four-body}):
\be
M_0 \approx 0.7 \, \eta_3^{1/4}  \left( \tan{\theta}\,M_\Theta M_{G'} \right)^{1/2}  ~~.
\label{eq:Mzero}
\ee
As mentioned in Section \ref{sec:phiWidth}, $\eta_3$ is a number that accounts for the 3-loop integrals, which are difficult to calculate; we expect $\eta_3$ to be of order 1, but values as low as 1/30 or so 
would not be very surprising.

\begin{figure}[b!]
\begin{center}
\includegraphics[width=0.95\textwidth]{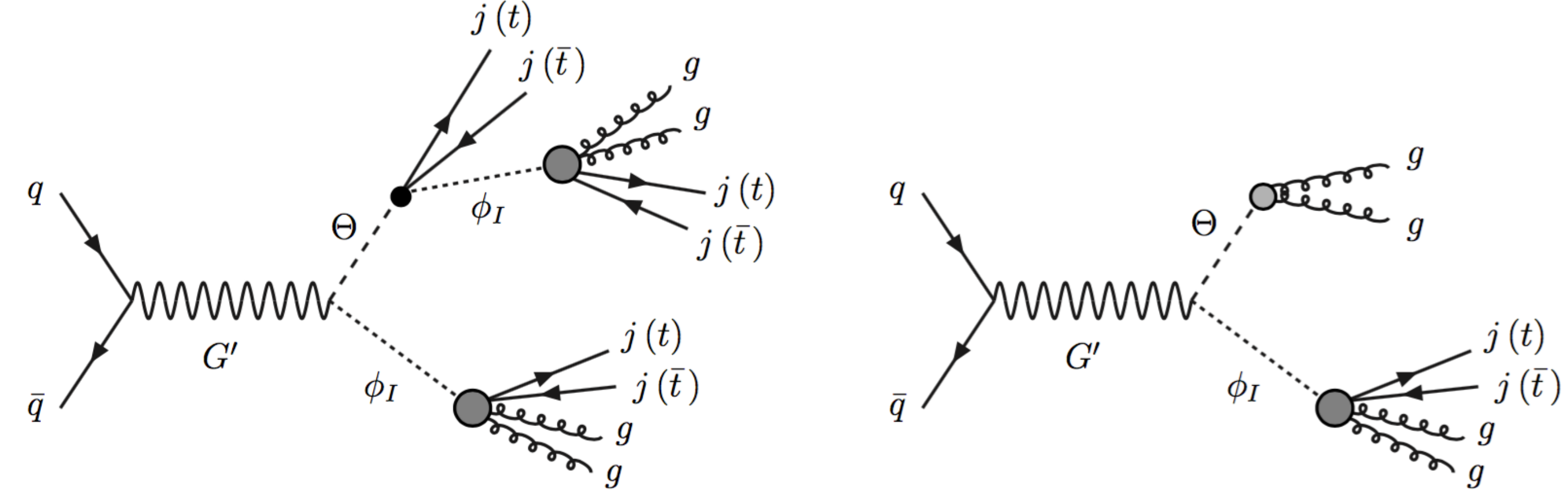}
\vspace*{-0.2cm}
\caption{Asymmetric $pp \to G' \to  \phi_I \Theta$ processes for $M_\Theta > M_{\phi_I}$, when 
the $\phi_I$ scalar decays into two gluons plus two quarks, which occurs at one-loop (see Figure \ref{fig:4body}).
Left diagram: scalar pair production is followed by a 3-body decay of the color-octet $\Theta$  into $\phi_I$ plus a quark-antiquark pair, which forms two jets, or less often a $t\bar t$ pair. 
Right diagram: $\Theta$ decays into two gluons, at one loop. }
\label{fig:HeavyTheta}
\end{center}
\end{figure}
%

\subsubsection{$\phi_I \to  gg q \bar q $}     

Let us first focus on the $M_{\phi_I} > M_0$ case, implying that $\phi_I$ decays predominantly into $gg q \bar q $.
An interesting process (see right diagram of Figure \ref{fig:HeavyTheta}) is $s$-channel coloron production followed by  the asymmetric coloron decay
 into $\Theta \, \phi_I$, and leading to a ten jet final state: $pp \rightarrow G' \to \Theta \phi_I \rightarrow  (\phi_I  q \bar q) \phi_I \to ((gg q \bar q) q \bar q) (ggq \bar q)  \rightarrow 10 j$.
Each parenthesis here denotes the presence of a resonance. 

Depending on the masses of the three ReCoM bosons, some of these jets may be merged. 
In particular, when $M_{G'}$ and $M_\Theta$ have 
the same order of magnitude while the scalar masses satisfy $M_\Theta \gg M_{\phi_I}$ and $M_{G'} - M_\Theta \gg M_{\phi_I}$,
the final state includes four jets with two of them having each a 4-prong substructure and an invariant mass given by $ M_{\phi_I}$.
We use the notation $J_{\phi_I}$ for such a jet  formed by the merger of two gluons and two quark jets arising from a $\phi_I$ decay.
The above cascade decay of the coloron is then $G' \to \Theta \phi_I \rightarrow  J_{\phi_I}J_{\phi_I} jj$.
If $M_{G'} - M_\Theta$ is near $M_{\phi_I}$ (a more tuned situation)
and $M_\Theta \gg M_{\phi_I}$, then the signal consists of seven jets, with only one of them 
being an $J_{\phi_I}$. 
For $M_{G'} \gg M_\Theta > M_{\phi_I}$, the whole $q\bar q \phi_I \to q\bar q (gg q' \bar q' )$ system (labelled by $J_{jj\phi_I}$) is boosted, 
and is approximately back-to-back against a $J_{\phi_I}$ (we will see though below that the branching fraction for $\Theta \to q\bar q \phi_I$ is small in this case).

There is also a $pp \rightarrow G' \to \Theta \phi_I \rightarrow (gg)(ggq\bar q) \to 6j$ process, where two gluon jets reconstruct the $\Theta$ mass, and the other four jets
reconstruct the $\phi_I$ mass. 
So far, no search in the six jet final state has been performed by imposing a separate mass constraint for two different resonances. 
Even though the intermediate states are different than in the process shown in Figure~\ref{fig:ThetaPhi} and discussed in Section~\ref{sec:octet-scalar}, the final state is identical.
Nevertheless, the range of parameters here is larger than in the case discussed in Section~\ref{sec:octet-scalar}, leading to novel boosted topologies. 
For $M_{G^\prime} > M_\Theta \gg M_{\phi_I}$ the resonant signal has three jets, with one of them being  an $J_{\phi_I}$,
and the other two jets forming a resonance at $M_\Theta$.  Requiring the total invariant mass of 
 three jets match the $G^\prime$ mass should provide a sensitive way to test these signatures. 
Using a simple estimate where the system of the four partons is boosted in the shape of a tetrahedron, we find that $\phi_I \rightarrow  ggq\bar q$
fits inside a cone of 
\be
\Delta R_{ggq\bar q} \simeq \frac{2 \sqrt{2} \,  M_{\phi_I}}{p_T(\phi_I)} \simeq \frac{4  \sqrt{2} \, M_{\phi_I}  }{M_{G^\prime}\left( 1 - M^2_{\Theta}/M^2_{G^\prime} \right) }~~.
\label{eq:DeltaR2}
\ee
Thus, one could  identify the $J_{\phi_I}$ jet using a normal jet-finding radius $R \sim 0.8$ for $M_{\phi_I} \approx 0.8$~TeV, $M_\Theta \approx 1$ TeV and $M_{G^\prime} \approx 6$~TeV. 

The quark-antiquark pair arising from either $\Theta \to q\bar q \phi_I$ or $\phi_I \to  gg q \bar q $ decays may be a $t\bar t$ pair, albeit the branching fraction is 
only 1/6 (or smaller if there is kinematic suppression), as opposed to 5/6 for light SM quarks.
In the case of  the  
$G' \to \Theta \phi_I \rightarrow  (\phi_I  q \bar q) \phi_I \to ((gg q \bar q) q \bar q) (ggq \bar q) $ cascade decay, the probability of having one of the quark-antiquark pairs to be 
$t\bar t$ can be as large as 34.7\%, and even the case where there are two $t\bar t$ pairs is potentially relevant, having  a probability of 6.9\% (without taking into account the $G'$ and $\Theta$ 
branching fractions).
Thus, if there are no large mass hierarchies, the signatures include $t\bar t + 8 j$ and $4t + 6j$. If $M_{\phi_I} \ll M_{G'}$, then the final states include $J_{\phi_I}$ jets with 4-prong substructure,
or boosted $t\bar t gg$ systems (we label them by $J_{t\bar t gg }$) with an invariant mass equal again to $M_{\phi_I} $. The latter have various interesting signatures; for example, when only one of the $W$ bosons produced by the top 
quarks decays leptonically, the boosted $t\bar t gg \to \ell b\bar b +4j +\met$ system will include a non-isolated lepton aligned with the missing energy.
Even more complicated objects occur when $M_{G'} \gg M_\Theta > M_{\phi_I} $. A boosted $\Theta \to t\bar t +4j$ (or less often  $\Theta \to 4t +jj$)  
system is then produced approximately  back-to-back with a $J^{}_{t\bar t gg}$.  

\begin{figure}[t]
\begin{center}
\vspace*{-0.1cm}
\includegraphics[width=\textwidth]{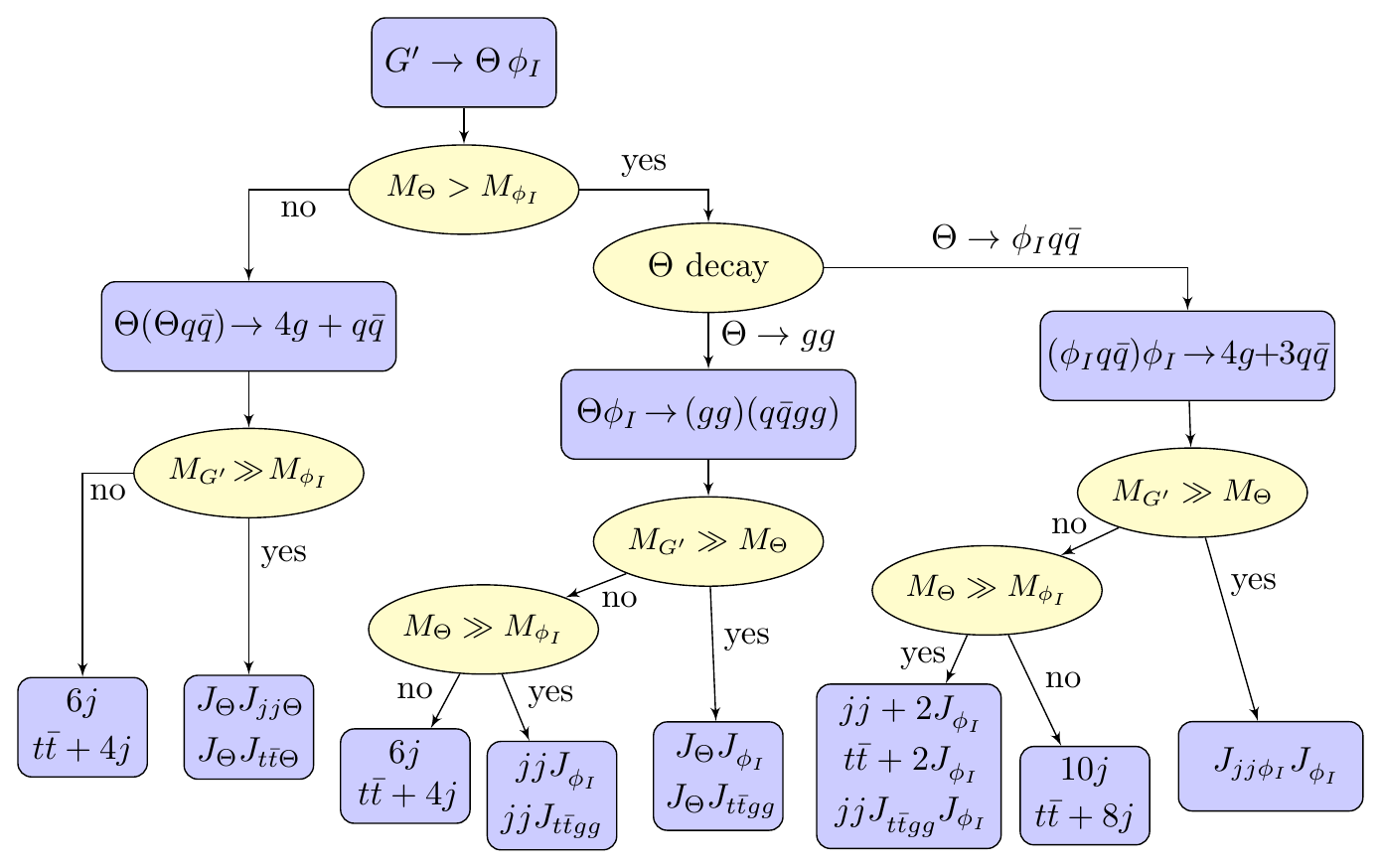}  
\vspace*{-0.4cm}
\caption{Signatures of the ReCoM from the asymmetric $pp \rightarrow G^\prime \rightarrow \Theta \phi_I$ production, and the corresponding 
mass relations. For $M_{\phi_I}  < M_\Theta$, only the 1-loop $\phi_I \to gg q\bar q$ decays are included here.
The jet  labelled by $J_\Theta$ has a 2-prong substructure, while $J_{jj \phi_I} $ has a 6-prong substructure; both have an invariant mass 
given by  $M_\Theta$.
The jets labelled by $J_{\phi_I}$ and $J_{jj\Theta}$ have a 4-prong substructure and an $M_{\phi_I}$ invariant mass. The boosted objects labelled by $J_{t\bar t gg }$ and $J_{t \bar t\Theta}$ have more complicated substructure and an $M_{\phi_I}$ invariant mass. 
For $M_{\phi_I} \lesssim O(1\,\mbox{TeV})$ and $M_{\phi_I}  < M_\Theta$, $\phi_I$ is so long-lived 
that $J_{\phi_I}$ and  $J_{t\bar t gg }$ may have a displaced origin.}  
\label{fig:flowchart1}
\end{center}
\end{figure}

The main LHC  signatures arising from the asymmetric $pp \to G' \to \Theta \, \phi_I$ process 
are summarized in Figure \ref{fig:flowchart1} (also including channels discussed in Section~\ref{sec:octet-scalar}). 
Note that some final states that require parameter tuning, or have smaller branching fractions, are not shown there.

For  $2 M_\Theta  > M_{G^\prime} > M_\Theta + M_{\phi_I}$,  the asymmetric channel $G' \to \Theta\phi_I$ is open and has a large branching fraction, while
the symmetric $G' \to \Theta\Theta$ channel is kinematically closed (even then, the QCD production of a pair of $\Theta$ scalars 
leads to many of the additional final states discussed in what follows).

\begin{figure}[t]
\begin{center}
\includegraphics[width=0.97\textwidth]{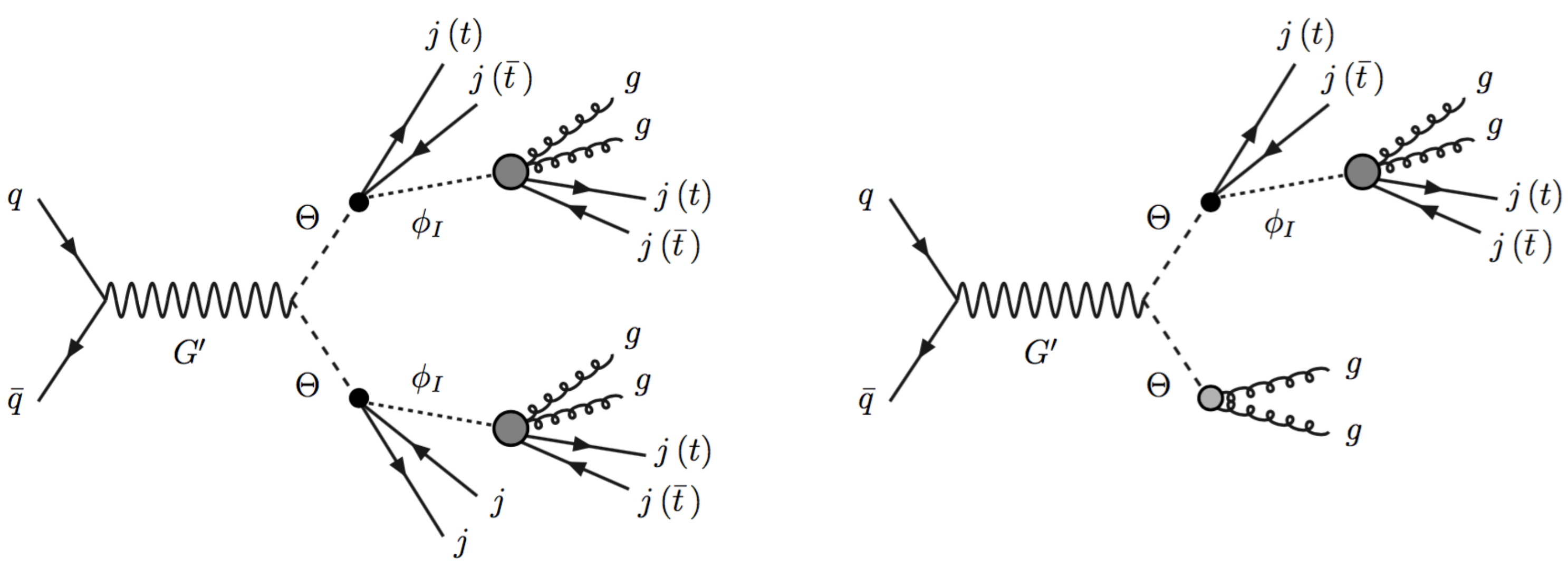}
\vspace*{-0.3cm}
\caption{Resonant production of a coloron that decays into a pair of color-octet scalars, for $M_\Theta > M_{\phi_I}$ when the singlet scalar  decays into two gluons plus two quarks. 
Left diagram: both $\Theta$ scalars undergo a cascade decay through an off-shell coloron.   
Right diagram: one $\Theta$ has a cascade decay, the other one decays at 1-loop into two gluons.  } 
\label{fig:ResonantCascade}
\end{center}
\end{figure}

For $M_{G^\prime} > 2 M_\Theta > M_\Theta + M_{\phi_I}$, both the asymmetric and the symmetric channels are open. 
Note that there is interference between the $pp \to G' \to \Theta\Theta$ and QCD $\Theta$ pair production.
The symmetric coloron decays lead to a variety of signatures in addition to those from Figure \ref{fig:flowchart1}.
Pair production of the color-octet scalar  followed by each $\Theta$ decaying into $jj\phi_I$ leads to a twelve jet final state:
$pp \rightarrow \Theta \Theta \rightarrow \phi_I \phi_I  + 4j \rightarrow 12 j$ (see  the left diagram of  Figure~\ref{fig:ResonantCascade}) if the main decay of $\phi_I$ is $\phi_I \rightarrow gg q \bar{q}$ at one loop. 
The searches for microscopic black holes via multi-jet final states \cite{CMS-PAS-EXO-15-007} 
may become sensitive to this ReCoM signal after the LHC accumulates enough luminosity. 
Specific to ReCoM, searching for a resonance in the invariant mass spectrum of all twelve jets can substantially improve the search sensitivity. 
 
For the parameter region with $M_\Theta \gg M_{\phi_I}$, the $\phi_I$ from $\Theta$ decays is likely to be boosted. Thus, the signature is 
$pp \rightarrow \Theta \Theta \rightarrow J_{\phi_I} J_{\phi_I} + 4j$. 
A possible search strategy is to require two jets to have a 4-prong substructure and approximately the same mass. In addition, one may try to reconstruct the 
$\Theta$ mass from the $C^2_4 = 6$ pairs of 3-jet resonances, similar to the searches for gluinos undergoing $R$ parity violating 
decays~\cite{Chatrchyan:2013gia,Aad:2015lea,ATLAS:2016nij}.

If the $\Theta \to \phi_I q \bar q$ and $\Theta \to gg$ decays have comparable branching fractions,
then the process $pp \rightarrow \Theta \Theta \rightarrow (\phi_I q\bar q) \,  (g g) \rightarrow 8 j$ (see  the right diagram of  Figure~\ref{fig:ResonantCascade}) has a large rate.
 For $M_\Theta \gg M_{\phi_I}$, the final state becomes $J_{\phi_I}+4j$. This  
 has the same combinatorial factor of 6 for reconstructing the intermediate $\Theta$ resonance. 

A substantial 3-body branching fraction for $\Theta$ requires that the mass ratio of $M_{\Theta}/M_{G^\prime}$ is not too small (see Figure~\ref{fig:theta-branching}). As a result, the 3-body decaying $\Theta$ is less likely to be boosted. 
To be more precise, for $M_{\phi_I} \ll M_\Theta \ll M_{G'}$ Eqs.~(\ref{eq:decaygg}) and (\ref{eq:3bodyTheta}) imply a ratio of branching fractions
\bear
\frac{B(\Theta \to \phi_I jj , \, \phi_I t\bar t \,  )}{ B(\Theta \to gg ) } &\simeq &  \frac{4 \pi \, \tan^2\! \theta }{405 \, \alpha_s \, (\pi^2/9 -1)^2 (1+r_{\mathcal{A}} )^2  }  \left( \frac{M_\Theta}{M_{G'} }\right)^2  \, 
\nonumber \\ [2mm]
& \approx & 0.1 \, \left(\frac{\tan\theta }{0.4}\right)^{\! 2}   \left( \frac{M_\Theta}{1 \; {\rm TeV} }\right)^{\! 2}  \left( \frac{5 \; {\rm TeV} }{M_{G'} }\right)^{\! 2}  ~~,
\label{eq:BRratio}
\eear 
where the 3-body branching fraction is summed over all six flavors and $r_{\mathcal{A}}\approx 0.53$.
The three jets produced in the $\Theta \to \phi_I q \bar q \to J_{\phi_I} jj $ decay fit inside an $\Delta R = 0.8$ cone provided 
$0.8 \gtrsim \sqrt{3} M_\Theta / p_T(\Theta) \simeq 2\sqrt{3}  \, M_\Theta / M_{G'}$, so that Eq.~(\ref{eq:BRratio}) implies an upper limit 
on the 3-body branching fraction, $B(\Theta \to  \phi_I jj , \, \phi_I t\bar t \,  ) \lesssim 12\%$ for $\tan\theta = 0.4$. Note that at this point in parameter space, 
which is almost optimal for a boosted $\Theta$ undergoing a 3-body decay, the branching fractions of $G'$ into scalars are only 
$B(G' \to \Theta\Theta) = 16\%$ and $B(G' \to \Theta\phi_I ) = 32\%$.     

\begin{figure}[t]
\begin{center}
\vspace*{-0.1cm}
\includegraphics[width=\textwidth]{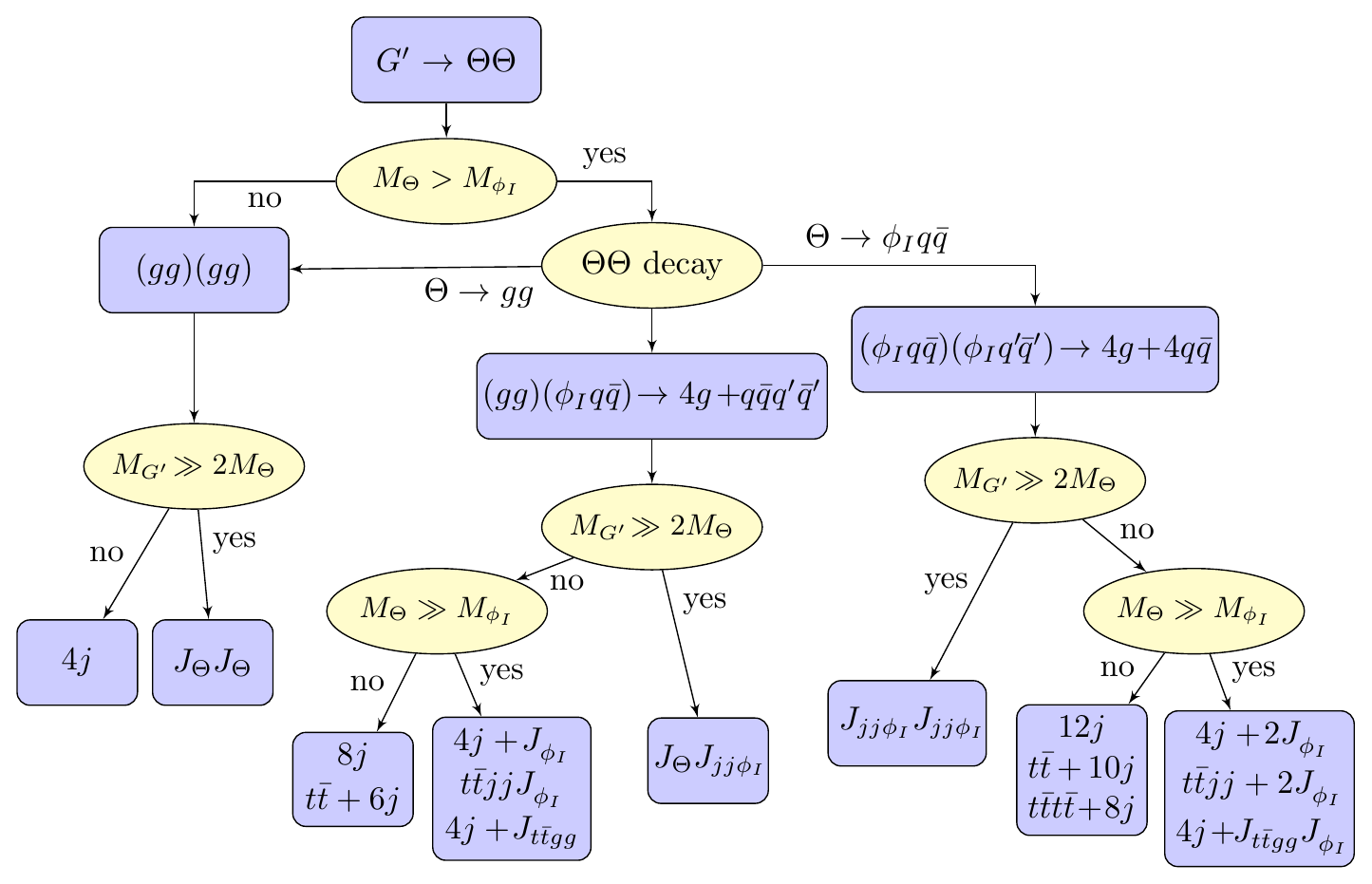}
\vspace*{-0.4cm}
\caption{Same as Figure~\ref{fig:flowchart1}, except that the final states shown here arise from 
 the $pp \rightarrow G^\prime \rightarrow \Theta \Theta$ process.} 
\label{fig:flowchart2}
\end{center}
\end{figure}

In the processes discussed above,  each of the quark-antiquark pairs from the $\Theta$ and $\phi_I$ decays 
may be a $t\bar t$ pair, with a branching fraction of 1/6 (for $M_\Theta - M_{\phi_I} \gg 2 m_t$ and $M_{\phi_I} \gg 2 m_t$).
Thus, the process where both $\Theta$'s undergo cascade decays (left diagram in Figure~\ref{fig:ResonantCascade})
includes a single $t\bar t$ pair with a probability of 38.6\%, and exactly two $t\bar t$ pairs with a probability of 11.6\%.

Pair production of $\Theta$ followed by both $\Theta$'s decaying to $gg$ has been discussed in Section~\ref{sec:octet-scalar}. 
The  main signatures of ReCoM arising from the $G' \to \Theta\Theta$ decays are summarized in Figure \ref{fig:flowchart2}.   
Certain processes with smaller branching fractions, such as $\Theta\Theta \to (\phi_I  t \bar t )( \phi_I  t \bar t) $, are not included there.

The $\phi_I \to gg q \bar q $ decays are either prompt or displaced depending on the values of the four parameters ($M_{\phi_I}$, $M_\Theta$, $M_{G'}$, $\tan\theta$) 
that enter the 4-body width given in Eqs.~(\ref{eq:phi-fourbody}) and (\ref{eq:CTheta}).  
The proper decay length of $\phi_I$ is approximately  proportional to 
$M_{G^\prime}^6M_\Theta^4 /M_{\phi_I}^{11}$, assuming that there is no fine tuning of the $M_\Theta/ M_{\phi_I} $ ratio. 
A $\phi_I$ produced in coloron decays has a boost $\gamma_{\phi_I} = E_{\phi_I} / M_{\phi_I} $, where $E_{\phi_I} $ is the energy of a $\phi_I$ in the lab frame. 
The decay length of $\phi_I$ in the lab frame is then
\bear
\lambda_4(\phi_I) & \approx &  
 \frac{663\,(4\pi)^4\,\tan^2\!{\theta}} {\alpha_s^5 (1+\tan^2\!{\theta})^4} \,   \, \frac{M_{G^\prime}^6\,M_{\Theta}^4 } {M_{\phi_I}^{12} } \, E_{\phi_I} 
\nonumber \\ [2mm]
&=&  O(0.03 \, {\rm cm}) \;  \left(\frac{\tan\theta }{0.3}\right)^{\! 2}  \left( \frac{700 \; {\rm GeV}}{M_{\phi_I}} \right)^{\! 12}     \left( \frac{M_{G^\prime}}{3\; {\rm TeV}} \right)^{\! 7} \left( \frac{M_\Theta}{1 \; {\rm TeV}} \right)^{\! 4}   \frac{E_{\phi_I}}{M_{G'}/6}   ~.
\label{eq:phi-length-4}
\eear
In the last line we assumed $\tan^2\!{\theta} \ll 1$. Note that the coloron is expected to be produced with small momentum, so the process
$pp \to G' \to  \Theta \Theta \rightarrow (\phi_I q\bar q)  \,  (\phi_I q\bar q)$ leads to 
a  $\phi_I$ energy  $E_{\phi_I} \sim O(M_{G^\prime}/6)$ for a typical event, while the process 
$pp \to G' \to  \Theta\phi_I \rightarrow (\phi_I q\bar q)  \,  \phi_I$ leads to 
a  slightly larger $E_{\phi_I}$.

The existing searches for displaced dijet resonances set an upper limit on the production cross section of a pair of displaced jets in the range of 1--10 fb  for  decay lengths larger than $O(1)$ mm and smaller than a meter or so \cite{CMS:2017oor} at the 13 TeV LHC.
As this topology occurs in the ReCoM for $M_{G'} \gtrsim 5$ TeV, for example from the $G'\to \Theta \phi_I \to jj J_\phi J_\phi$ process with displaced $J_\phi$,
the production cross section shown in Figure~\ref{fig:prodthetaphi} is not yet constrained.

\bigskip

\subsubsection{$\phi_I \to W^+W^-$, $\gamma Z$, $ZZ$}  

When $\phi_I$ is lighter than both $\Theta$ and the scale $M_0$ introduced in Eq.~(\ref{eq:Mzero}),
its leading decays are into two electroweak gauge bosons (with the exception of two photons) and occur at three loops, as in the right-hand diagram in Figure~\ref{fig:2loops}.
To find the LHC signatures, one can repeat the previous discussion with the 4-body decay $\phi_I \rightarrow gg q\bar{q}$ replaced by the 2-body ones. 
The branching fractions of $\phi_I\rightarrow W^+W^-, \gamma Z, \, ZZ$ are 65\%, 19\%, $16\%$, respectively, or smaller 
if the $\phi_I \rightarrow gg q\bar{q}$ width is not negligible.

\begin{figure}[t!]
\begin{center}
\includegraphics[width=0.94\textwidth]{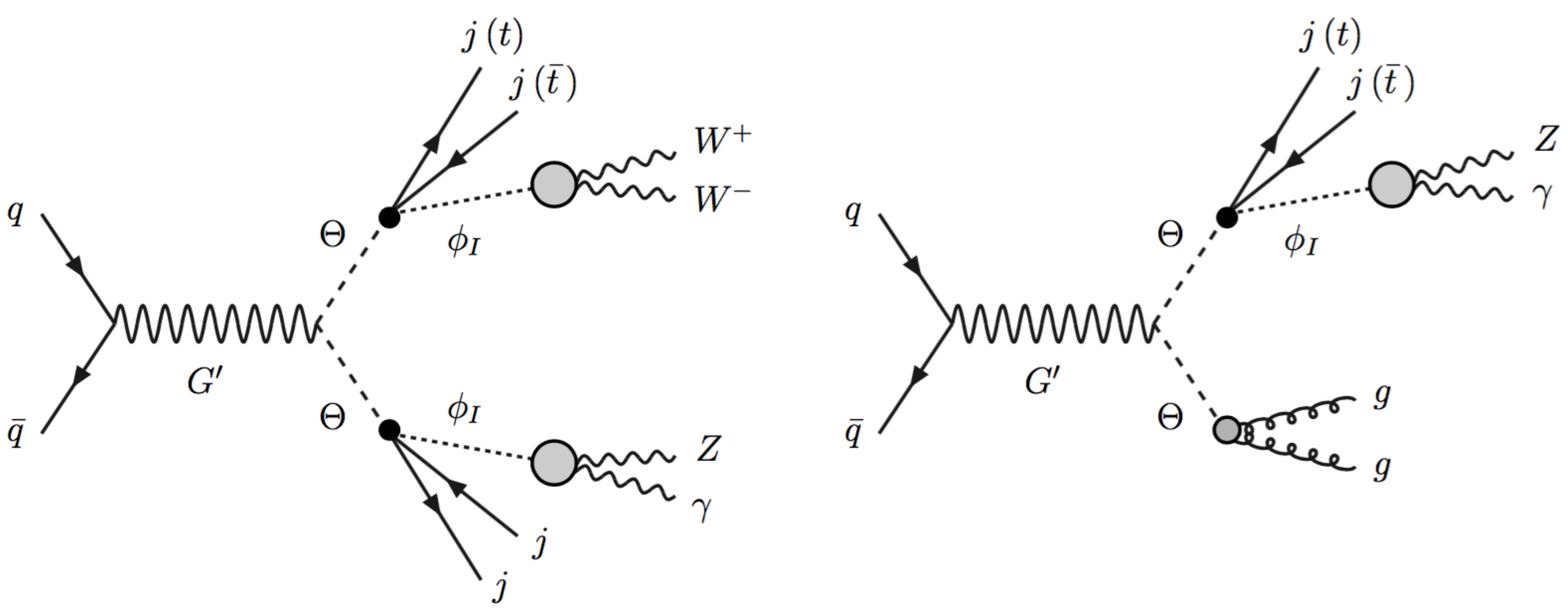}
\vspace*{-0.3cm}
\caption{Representative LHC processes from symmetric coloron decays, for $M_\Theta > M_{\phi_I}$ when $\phi_I$ decays into pairs of electroweak bosons at
three loops (represented by the largest gray disc). The $\gamma Z$ and  $W^+W^-$ systems are likely to be boosted and to originate from displaced vertices.}
\label{fig:diagramsWW}
\end{center}
\end{figure}

The signatures of 
asymmetric coloron decays  are $G' \to \Theta \phi_I  \to  (\phi_I \, q\bar q / t \bar t ) \phi_I  \to 2(\gamma Z/ZZ/W^+W^-) +jj /t\bar t$ when  $\Theta$ has 3-body decays, 
and $\Theta \phi_I  \to  (\gamma Z/ZZ/W^+W^-) +jj $ in the case of  $\Theta \to gg$.
Pair production of the color-octet scalars, through QCD and $s$-channel $G'$,
leads to  $\Theta \Theta \rightarrow (\phi_I q\bar q)(\phi_I q \bar q)  \rightarrow 2(\gamma Z/ZZ/W^+W^-) + 4j$ and 
the same with one or two $q\bar q$ pairs replaced by $t\bar t$ pairs, as well as
$\Theta \Theta \rightarrow (\phi_I q\bar q) \,  (g g) \rightarrow (\gamma Z/ZZ/W^+W^-) + 4j$, as shown for example in Figure~\ref{fig:diagramsWW}. 
Note that the boson pairs tend to be boosted given that  $M_{\phi_I} \lesssim 1$ TeV in this case, while the coloron has a mass of several TeV. We will use the notation $J_{WW}$, $J_{\gamma Z}$ and $J_{ZZ}$
for a boosted di-boson system of invariant mass equal to $M_{\phi_I}$, independent of how the $W$ or $Z$ bosons decay.

For hadronic decays of $ZZ$ and $W^+W^-$, $\phi_I$ could still behave as a 4-prong jet with two internal $Z$ or $W$-jets. 
For the final states with a photon or leptonic decays of $Z$ and $W$ bosons, the collider signatures contain $\gamma \, \ell^+ \ell^-  + \mbox{jets}$, $\gamma + \met + \mbox{jets} $, 
$\ell^\pm + \met + \mbox{jets} $, $\ell^+ \ell^- + \met + \mbox{jets} $, or more leptons when there are up to four  weak bosons from two $\phi_I$ decays. 

The  2-body partial widths of $\phi_I$, given in Eq.~(\ref{eq:decay-two-boson}), 
imply that  the lab-frame decay length of $\phi_I$ in this case is
\bear
\lambda_2(\phi_I) & \approx &  
 \frac{4\,(4\pi)^6\,c_W^4\,s_W^4} {25\,\eta_3^2\,\alpha^2\alpha_s^5 \tan^2\!{\theta} (1+\tan^2\!{\theta})^4\,(3 c_W^4 + s_W^4) }  \, \frac{M_{G^\prime}^2 } {M_{\phi_I}^{4} } \, E_{\phi_I} 
\nonumber \\ [2mm]
&=&  O(0.03\, {\rm cm}) \;  \left(\frac{\tan\theta }{0.3}\right)^{\! 2}  \left( \frac{700 \; {\rm GeV}}{ \sqrt{\eta_3} \, M_{\phi_I}} \right)^{\! 4}     \left( \frac{M_{G^\prime}}{3\; {\rm TeV}} \right)^{\! 3} \frac{E_{\phi_I}}{M_{G'}/6}   ~.
\label{eq:phi-length-2}
\eear
Note the milder dependence on $M_{\phi_I}$ compared to Eq.~(\ref{eq:phi-length-4}), as well as the dependence on the numerical coefficient $\eta_3$.
In most cases the $\phi_I \to W^+W^-, \gamma Z, \, ZZ$ decays are associated with 
displaced vertices.  For $\eta_3$ of order one, the displacement varies between the millimeter scale (for $\phi_I \gtrsim 500$ GeV) and the meter scale
(for $\phi_I \approx 100$ GeV), while for $\eta_3$ of order 0.1 the displacement grows by two orders of magnitude.
Thus, we expect  $WW$, $\gamma Z$ and $ZZ$ resonances which are both boosted and displaced.
So far there have been no dedicated collider searches for boosted systems with displaced vertices ({\it e.g.}, current searches for long-lived gluinos~\cite{Khachatryan:2016sfv,Aaboud:2017iio} do not cover boosted signatures).

 The $\phi_I \to \gamma Z$ decay may in principle be constrained by existing searches with non-pointing photons without special requirement of jet multiplicity~\cite{Aad:2014gfa}.  For the lifetime of the long-lived particle in the range of 250~ps to about 100~ns, the upper limit  on the signal production cross section at the 8 TeV LHC ranges from 1~fb to 100~fb. 
By comparing to the 13 TeV  LHC cross section for the production of a $\Theta$ pair shown in Figure~\ref{fig:prodThetaTheta}, and multiplying by the 
branching fractions for one of the $\Theta$'s  to decay into $\phi q \bar q$ and also for $\phi_I \to \gamma Z$, we derive a lower limit on $M_\Theta$ of roughly 0.7 TeV
when the coloron is not too heavy [so that the 3-body $\phi$ branching fraction is not too suppressed, see Eq.~(\ref{eq:BRratio})].
The signal sensitivity can be improved by also requiring the presence of high $p_T$ jets. Therefore, for a $\phi_I$ mass below about 700 GeV, searches with a displaced $J_{\gamma Z}$ 
and prompt high-$p_T$  jets could lead to a clean discovery.  

When $M_{\phi_I} < M_Z$, which is a natural region of parameter space given that $\phi_I$ is a pseudo-Nambu Goldstone boson,
the main decays  of $\phi_I$ are 3-loop 3-body processes, which include a photon and proceed via an off-shell $Z$ boson: $\phi_I \to \gamma jj, \, \gamma \nu\bar\nu, \, \gamma \ell^+\ell^-$.
The decay length is then longer than the detector, so $\phi_I$ behaves as a missing particle in collider experiments. The collider signatures of coloron decays then include missing transverse energy 
plus prompt jets or top quark pairs when there are 3-body $\Theta$ decays, or plus a dijet resonance when one of the $\Theta$ decays into two gluons. 

\bigskip \bigskip      

\section{Discussion and conclusions
} \setcounter{equation}{0}
\label{sec:conclusions}

The ReCoM is a UV-complete gauge extension of QCD. The sector that spontaneously 
breaks the $SU(3)_1\times SU(3)_2$ gauge symmetry includes a color-octet scalar 
($\Theta$) and two singlet scalars ($\phi_I$ and $\phi_R$).  It is natural that $\Theta$ and $\phi_I$ are lighter than the coloron, as they are 
pNGB's associated with an $SO(18) \to SO(17)$ global symmetry breaking. In addition, $\phi_I$ is the pNGB associated with the breaking of a global $U(1)_\Sigma$ symmetry.

All the SM quarks are triplets under one (the first, by convention)  of the $SU(3)$ gauge groups, so the coloron has 
flavor-universal couplings. The ReCoM belongs to a small but remarkable class of gauge extensions 
of the SM that does not require anomalons, {\it i.e.}, fermions beyond the SM quarks and 
leptons that cancel the gauge anomalies.\footnote{Another gauge extension of QCD without anomalons 
is  $SU(3)_1\times SU(3)_2 \times SU(3)_3$ with each generation of SM quarks charged under a different $SU(3)$ (see also \cite{Agrawal:2017evu} for a possible solution to the strong $\mathcal{CP}$ problem based on this product of gauge groups).} 

As there are no significant constraints on the coloron from flavor-changing processes\footnote{Flavor constraints
are stringent  \cite{Chivukula:2010fk}   in models where the SM quarks transform under different 
$SU(3)_1\times SU(3)_2$ representations.} or electroweak  observables, the only relevant limits on the 
coloron mass ($M_{G'}$) and coupling ($\tan\theta$) are set by
searches at the LHC.  Dijet resonance searches set the limits in the ($\tan\theta$, $M_{G'}$) plane 
shown in Figure~\ref{fig:theta-exclusion}. 
The $t\bar t$ resonance searches are weaker because the coloron branching fraction into $t\bar t$ is smaller than 
the $jj$ one by a factor of 5.

For $\tan\theta \lesssim 0.45$, the dominant coloron decays are into a pair of scalars. 
The $G'\to  \Theta \, \phi_I$ and $\Theta\Theta$ decays have typically the largest branching fractions 
(unless the masses are near the kinematic threshold, see Figure~\ref{fig:Gprime-branching}).
The subsequent decays of $ \Theta$ and $\phi_I$ lead to a variety of final states, depending on four parameters: 
$M_{G'}$, $M_\Theta$, $M_{\phi_I}$ and $\tan\theta$.
When $M_{\phi_I} > M_\Theta$, the color octet $\Theta$ decays into a pair of gluons, and $\phi_I$ decays into $jj \Theta$,
so that the final state includes up to six jets,  with some of them merged depending on how light the scalars are compared to the coloron.

When $M_{\phi_I} < M_\Theta$, the main decays of $\Theta$ are into $q\bar q\phi_I$ (where $q$ is a SM quark jet), $t \bar t \phi_I$ or $gg$, while the decays of $\phi_I$ have 
highly suppressed widths. The latter include 3-loop 2-body decays (into $W^+W^-$, $\gamma Z$ or $ZZ$), 
as well as 1-loop 4-body decays (into $gg q\bar q$ or $gg \, \bar t t$). The ratio of these widths is estimated in Eq.~(\ref{eq:two-body-over-four-body}),
where the parameter $\eta_3$ introduced in  Eq.~(\ref{eq:3loopOps}) accounts for the uncertainty arising from the 3-loop integrals (computing these is beyond the scope of this paper).
We expect that $\eta_3^2$ lies between $O(1)$ and $O(10^{-3})$, but even this large uncertainty has a relatively small impact on the 
possible final states due to the dependence on high powers of the masses ({\it e.g.}, the ratio of widths is proportional to $M_{\phi_I}^8$).
For $M_{\phi_I}$ larger than the mass scale $M_0$ [see Eq.~(\ref{eq:Mzero})], which is of the order of 1 TeV, 
the main decay is  $\phi_I \to q\bar q gg$. Although this may be a prompt decay for larger values of  $M_{\phi_I}$, it 
may also lead to displaced vertices [see Eq.~(\ref{eq:phi-length-4})].  Final states with top quarks have 
smaller branching fractions (in some cases only by a factor of  5/4 or 5/3, if there are four or three quark-antiquark pairs)  but are still promising.
The final states with multiple jets or top quarks are shown in Table~\ref{table:1}. For $M_{\phi_I} \ll M_{G'}$, the $gg q \bar q$  system is boosted 
and forms a jet with 4-prong substructure (labelled by $J_{\phi_I}$), while the boosted $gg t\bar t$ system is even more complicated. 

For $M_{\phi_I}$ lighter than the mass scale $M_0$,  the main decay is into a pair of electroweak bosons ($W^+W^-$, $\gamma Z$, $ZZ$), and the 
decay length of $\phi_I$ varies between a fraction of millimeter to longer than  1 cm  [see Eq.~(\ref{eq:phi-length-2})]. For $M_{\phi_I} < M_Z$,
the main decay is a 3-loop 3-body decay into a photon and an off-shell $Z$ boson; in that case the decay is likely to be outside the detector, 
so $\phi_I$  would appear as missing transverse energy. 
 The channels with electroweak bosons that have  large branching fractions, and which do not rely on fine tuning of masses, are shown in Table~\ref{table:2}. 

So far there have been dedicated searches for very few of the final states listed in Tables~\ref{table:1} and \ref{table:2}. Even though some 
of these final states can be lumped in certain multi-jet or multi-lepton searches, it is important 
for the ATLAS, CMS and LHCb collaborations to perform dedicated searches for each of these final states.
The sensitivity of a dedicated search far exceeds that of a generic search, especially when the signature involves nested resonances,
multi-prong jet substructure, non-isolated leptons or photons, or displaced vertices. Furthermore, depending on the parameter values and the improvements in experimental 
techniques, any of these signatures may become a discovery mode.

Some of these final states are also encountered in other models. For example,  the dominant signals 
in the ReCoM for a range of masses are $J_{\phi_I} \, jj $ or $J_{\phi_I} +4j $,  with a narrow $J_{\phi_I}$. These would appear as a  $3j$ or $5j$ 
resonance, respectively, which so far have not been searched for. The former is also predicted in models where a 
vector-like quark is produced at 1-loop in the $s$-channel, from a gluon-quark initial state, and decays via a dimension-6 operator into three SM quarks. 

Even dijet resonance searches, if augmented by substructure 
techniques, may reveal unusual final states, such as   $J_{jj \Theta} J_\Theta$,  in which one jet has a 
4-prong substructure and the other jet has a 2-prong substructure and a smaller mass.
Other final states are more peculiar, such as those 
with a large number of electroweak bosons ({\it e.g.}, $ \gamma Z W^+W^-  + 4j $ or $W^+W^-W^+W^- t\bar t jj $),
or those involving a displaced $\gamma Z$ resonance.
Clearly, there are many new opportunities  for new physics searches at the LHC.

\bigskip\bigskip\bigskip\bigskip

{\it  Acknowledgments:}  {\small We thank  Sida Lu, Nhan Tran and Qianfei Xiang for useful discussions and comments. 
The work of YB is supported by the U. S. Department of Energy under the contract No. DE-SC-0017647. The work of BD has been supported by Fermi Research Alliance, LLC under Contract No. DE-AC02-07CH11359 with the U.S. Department of Energy, Office of Science, Office of High Energy Physics. 
}
\begin{table}[h!]
\begin{center}
\renewcommand{\arraystretch}{1.25}
\begin{tabular}{|c|c|c|}\hline 
partonic process from $G'$ decay &  final state   &  mass relation 
 \\ \hline \hline
         &          $6j$           &  \hspace*{-2mm}  $ M_{G'}  > M_{\phi_I} > M_\Theta$  \hspace*{-2mm}  \\  [-.85em]
$ \Theta\phi_I \to \Theta (\Theta q \bar q) \to (gg) \, ((gg) q \bar q) $              &                       &    \\  [-.85em]
         &   $J_\Theta J_{ jj\Theta}$  &      $M_{G'} \gg M_{\phi_I} > M_\Theta $                \\ [0.1em]     
\hline
  &  $10j$   &   $M_{G'} > M_\Theta > M_{\phi_I}$       \\ [-.2em]  
$\Theta\phi_I \to (\phi_I q \bar q) \phi_I  \to ((gg q\bar q) q \bar q)(gg q\bar q) $               &   $J_{\phi_I} + 6 j  $       &  $M_{G'} - M_\Theta \gg  M_{\phi_I}$    \\ [-.1em]
       &   $J_{\phi_I}J_{\phi_I} \,  j j  $                 &    $M_{G'} > M_\Theta \gg  M_{\phi_I}$      \\ [0.1em]
 \hline
   &  $6 j $    &   $M_{G'} > M_\Theta >  M_{\phi_I}$       \\ [-.2em]
$\Theta\phi_I \to (gg)(gg q \bar q)$               &    $J_{\phi_I} \, jj $    &   $M_{G'} - M_\Theta \gg  M_{\phi_I}$     \\ [-.1em]
     & $J_{\Theta}J_{\phi_I}$        &  $M_{G'} \gg M_\Theta >  M_{\phi_I}$   \\ [.1em]
 \hline
      &         $12j$          &   $M_\Theta >  M_{\phi_I}$   \\ [-.8em]
$\!\!   \Theta\Theta \!\to\!   (\phi_I  q \bar q)  \, (\phi_I  q \bar q )  \!\to  ((gg q \bar q) q \bar q)  ((gg q \bar q) q \bar q)   \!\!   $    &  &     \\ [-.8em]
  &    $J_{\phi_I}J_{\phi_I} + 4j$    &    $M_\Theta \gg  M_{\phi_I}$  
  \\ \hline
      &  $8j$   &   \hspace*{-2.5mm}  $M_\Theta > M_{\phi_I} $  \hspace*{-2.5mm}    \\ [-.85em]
$\Theta\Theta \to   (\phi_I  q \bar q)  \, (gg)  \to ((gg q \bar q) q \bar q)  \,  (gg) $    &    &       \\ [-.85em]
 &       $J_{\phi_I} + 4j$          &    \hspace*{-2.5mm}  $M_\Theta \gg  M_{\phi_I} $  \hspace*{-2.5mm} 
\\ \hline
      &        $4j$           &     
      \\ [-.9em]
$\Theta\Theta \to (gg)(gg)  $                &  &       \\ [-.85em]
 & $J_\Theta J_\Theta$ &       $M_{G'} \gg 2 M_\Theta$
\\ \hline  \hline
  &    $t\bar t + 8j$     &   $M_{G'} > M_\Theta >  M_{\phi_I}$       \\ [-.85em]  
$\Theta\phi_I \to (\phi_I q \bar q) \phi_I  \to  (gg t \bar t)  \, (gg q \bar q)  \,  q  \bar q $               &   &    \\ [-.85em] 
       &   $ J_{gg t \bar t} \, J_{\phi_I} \,  jj  $                &   $M_{G'} > M_\Theta \gg  M_{\phi_I}$    \\ [0.1em]
 \hline
       &      $ t\bar t + 10j $          &     \hspace*{-3mm} $ M_\Theta >  M_{\phi_I}$  \hspace*{-3mm}  \\ [-.2em]  \cline{2-3} 
$\Theta\Theta \to  (\phi_I \,  t \bar t /q \bar q )  \, (\phi_I  q \bar q)  \to  (gg q \bar q  t \bar t)  \,  ((gg q \bar q)  q \bar q) $   &   $\!  J_{gg t \bar t} \, J_{\phi_I} \! +4j \!  $    &      \\ [-.9em]
& &    $M_\Theta \gg  M_{\phi_I}$        \\ [-.75em]
 &  $ t \bar t  \,J_{\phi_I} J_{\phi_I} \,  jj $   &  
 \\ \hline
        &      $ t\bar t t\bar t + 8j $          &     \hspace*{-3mm} $M_\Theta >  M_{\phi_I}$  \hspace*{-3mm}  \\  [-.2em]  \cline{2-3}  
        &      $ t\bar t   J_{gg t \bar t} \, J_{\phi_I} \,  jj $          &     \\ [-.85em]
$\Theta\Theta \to  (\phi_I  t \bar t /q \bar q )  \, (\phi_I   t \bar t /q \bar q)  \to  (gg q \bar q  t \bar t)  \,  (gg q \bar q  t \bar t) $   &     &     \\ [-.8em]
& $J_{gg t \bar t} J_{gg t \bar t} \!+\!4j \! $  
  &  $M_\Theta \gg  M_{\phi_I}$    \\  
 &  $ t \bar t   t \bar t   \,J_{\phi_I} J_{\phi_I}  $   &   
   \\ \hline
        &      $ t\bar t + 6j $          &    \hspace*{-3mm}    $M_\Theta >  M_{\phi_I}$   \hspace*{-3mm}  \\  [-.2em]  \cline{2-3}
$\Theta\Theta \to  (\phi_I \,  t \bar t / q \bar q )  \, (g g)   \to   (gg q \bar q t \bar t)  (gg) $  &   $J_{gg t \bar t} +4j $ &    \\ [-.9em]
&  &  $M_\Theta \gg  M_{\phi_I}$    \\ [-.75em]
 &     $ t \bar t  \,J_{\phi_I} \,  jj $ &  
 \\ \hline
\end{tabular}
\caption{Main LHC processes predicted in the ReCoM when $M_{\phi_I} > M_\Theta$, or when  $M_{\phi_I} < M_\Theta$ with $\phi_I$ decaying to $ggq\bar q$. Each parenthesis represents a resonance. 
$J_{\phi_I} $ is a jet with 4-prong 
substructure arising from a boosted $\phi_I \to ggq\bar q$, 
and $J_{g g t \bar t } $ is a  boosted $\phi_I \to gg t\bar t $.
Depending on the parameters from Eq.~(\ref{eq:phi-length-4}), 
$J_{\phi_I} $ and $J_{g g t \bar t } $ may have a displaced vertex for $M_{\phi_I} \lesssim$ 1 TeV. 
$J_\Theta$ is a jet with 2-prong substructure due to a  boosted $\Theta \to gg$.  
$J_{jj \Theta}$ is a jet with 4-prong substructure due to a boosted $\phi_I$. 
Additional channels  are less likely to occur ({\it e.g.,}   $\Theta\Theta  
\to  J_{jj\phi_I}  J_{\Theta}$, or $\Theta\phi_I \to 4t + 6j $).   }
\label{table:1}
\end{center}
\end{table}

\vfil\newpage

\begin{table}[h!]
\begin{center}
 \vspace{-0.2cm}
\renewcommand{\arraystretch}{1.25}
\begin{tabular}{|c|c|c|}\hline 
partonic process from $G'$ decay &  final state   &  mass relation 
 \\ \hline \hline
   &  $\!  W^+W^-W^+W^- \! jj \!$    &   $M_{G'} > M_\Theta >  M_{\phi_I}$       \\ [.1em]   
$\Theta\phi_I \to (\phi_I q \bar q) \phi_I  \to ((WW) q \bar q)(WW) $           &  $ W^+W^- J_{_{WW}} jj $    &   $M_{G'} - M_\Theta \gg  M_{\phi_I}$       \\ [.1em]  
  &  $J_{_{WW}} J_{_{WW}}  \,  j j $    &   $M_{G'} > M_\Theta \gg  M_{\phi_I}$       \\ [.1em]  
 \hline
   &  $ \gamma Z W^+W^- jj $    &    $ M_\Theta >  M_{\phi_I}$          \\ [.1em]  
      &  $ \gamma Z J_{_{WW}} jj $    &  $M_{G'} - M_\Theta \gg  M_{\phi_I}$        \\ [-.4em]  
$\Theta\phi_I \to (\phi_I q \bar q) \phi_I  \to (WW)(\gamma Z)  q \bar q $               &   &    \\ [-1.0em]
     &  $ W^+W^- J_{_{\gamma Z}} \, jj $    &   $M_{G'} - M_\Theta \gg  M_{\phi_I}$         \\ [.1em]  
  &  $J_{_{\gamma Z}} J_{_{WW}}  \,  j j $    &   $M_\Theta \gg  M_{\phi_I}$       \\ [.1em]  
 \hline
   &  $W^+W^- jj$    &   $ M_\Theta >  M_{\phi_I}$       \\ [.0em]
$\Theta\phi_I \to (gg)(WW)$       &  $J_{_{WW}} \, jj $    &   $M_{G'} - M_\Theta \gg  M_{\phi_I}$       \\ [.1em]
      & $J_{_{WW}} J_{\Theta}$        &  $M_{G'} \gg M_\Theta >  M_{\phi_I}$   \\ [.1em]
 \hline
   &  $ \gamma Z jj$    &   $ M_\Theta >  M_{\phi_I}$       \\ [.0em]
$\Theta\phi_I \to (gg)( \gamma Z)$       &  $J_{_{ \gamma Z}} \, jj $    &   $M_{G'} - M_\Theta \gg  M_{\phi_I}$       \\ [.1em]
      & $J_{_{ \gamma Z}} J_{\Theta}$        &  $M_{G'} \gg M_\Theta >  M_{\phi_I}$   \\ [.1em]
 \hline
      &         $4\,W + 4j $          &       $M_\Theta >  M_{\phi_I}$      \\ [-.75em]
      $\!\!\!   \Theta\Theta \!\to\!   (\phi_I  q \bar q)  \, (\phi_I  q \bar q )  \!\to  ((WW) q \bar q)  ((WW) q \bar q)   \!\!\!  $    &  &     \\ [-.75em]
       &  $J_{_{WW}} J_{_{WW}} + 4j$  &    $M_\Theta \gg  M_{\phi_I}$  
  \\ \hline
        &         $ \gamma Z W^+W^-  + 4j $          &       $M_\Theta >  M_{\phi_I}$        \\ [-.75em]
      $\!\!   \Theta\Theta \!\to\!   (\phi_I  q \bar q)  \, (\phi_I  q \bar q )  \!\to  ((\gamma Z) q \bar q)  ((WW) q \bar q)   \!\!   $    &  &     \\ [-.75em]
       &  $J_{_{\gamma Z }} J_{_{WW}} + 4j$  &    $M_\Theta \gg  M_{\phi_I}$   
  \\ \hline
      &      $W^+W^-  + 4j$          &     $M_\Theta >  M_{\phi_I}$    \\ [-.8em]
$\Theta\Theta \to   (\phi_I  q \bar q)  \, (gg)  \to ((WW) q \bar q)  \,  (gg) $    &    &       \\ [-.8em]
     &      $J_{_{WW}}  + 4j$          &    \hspace*{-2.5mm}  $ M_\Theta \gg  M_{\phi_I} $  \hspace*{-2.5mm} 
 \\ \hline
      &      $\gamma Z  + 4j$          &     $M_\Theta >  M_{\phi_I}$     \\ [-.8em]
$\Theta\Theta \to   (\phi_I  q \bar q)  \, (gg)  \to ((\gamma Z) q \bar q)  \,  (gg) $    &    &       \\ [-.8em]
     &      $J_{_{\gamma Z}}  + 4j$          &    \hspace*{-2.5mm}  $M_\Theta \gg  M_{\phi_I} $  \hspace*{-2.5mm} 
\\ \hline  \hline
       &      $4\,W +  t\bar t \, jj $          &        $M_\Theta >  M_{\phi_I}$       \\ [-.7em]
  $  \!\Theta\Theta \to  (\phi_I  t \bar t)  \, (\phi_I  q \bar q)  \to  ((WW) t \bar t)  \,  ((WW)  q \bar q)  \! $  & &  \\ [-.7em]
  & $ J_{WW} J_{WW} \, t \bar t  jj $   &  $ M_\Theta \gg  M_{\phi_I}$  
 \\ \hline
        &      $\gamma Z W^+W^- t\bar t \, jj $          &      $M_\Theta >  M_{\phi_I}$            \\ [-.7em]
$\Theta\Theta  \to  (\phi_I  t \bar t)  \, (\phi_I  q \bar q)  \to  (\gamma Z) (WW) \,  t \bar t \,  q \bar q $  &  & \\ [-.7em]
        & $ J_{\gamma Z} J_{_{WW}} \, t \bar t \,  jj $   &  $M_\Theta \gg  M_{\phi_I}$   
 \\ \hline
\end{tabular}
\caption{LHC processes predicted in the ReCoM when $M_{\phi_I} \! < \! M_\Theta$ and $\phi_I \! \to \! WW$ or $\gamma Z$.  
$J_{_{WW}} $ and $J_{\gamma Z} $ represent  a  $W^+W^-$ or  $\gamma Z$ system produced in the decay of a 
boosted $\phi_I$. These originate from a displaced vertex for $M_{\phi_I} \lesssim$ 1 TeV, and may appear as $\met$ for $M_{\phi_I} \lesssim$ 100 GeV. 
Additional channels  ({\it e.g.,} those involving  two $t\bar t$ pairs or  $\phi_I  \to ZZ$)
have smaller branching fractions. The mass relations displayed here are necessary but not sufficient ({\it e.g.,} $ \gamma Z J_{_{WW}} jj $ also requires that $M_\Theta$
is not much larger than $M_{\phi_I} $).
}
\label{table:2}
\end{center}
\end{table}

\bigskip\bigskip

\section*{Appendix A: Ratios of scalar masses}
\renewcommand{\theequation}{A.\arabic{equation}}
\addcontentsline{toc}{section}{\normalsize Appendix A: Ratios of scalar masses}
 \setcounter{equation}{0}
\label{app:mass-ratios}

We now derive the allowed range for the scalar mass ratio $M_{\phi_I}/M_\Theta$. 
As mentioned in Section~2, in the limit $\mu_\Sigma \to 0$, the potential has a global $U(1)_\Sigma$ symmetry, which is broken by the $\Sigma$ VEV.
The corresponding Nambu-Goldstone is $\phi_I$, so that $M_{\phi_I}/M_\Theta \to 0 $ for $\mu_\Sigma \to 0$.

The upper limit of $M_{\phi_I}/M_\Theta$ is more complicated to derive. Let us first 
use Eq.~(\ref{eq:scalar-masses}) to write
\be
\frac{M_{\phi_I}}{M_\Theta}  =  \sqrt{3} \,  \left(  2 +  \frac{\sqrt{2} \, \kappa \, f_\Sigma}{\sqrt{3} \, \mu_\Sigma}  \right)^{\! -1/2}  ~~.
\ee
The necessary and sufficient condition for the potential $V(\Sigma)$ [see Eq.~(\ref{eq:sigma-pot})] to be bounded from below, derived in \cite{Bai:2017zhj}, is 
\be
 \kappa >   {\rm max}  \left\{ - \lambda , - 3 \lambda \right\}  ~~.
 \label{eq:bounded}
\ee
Imposing this condition, and using the expression of the VEV value $f_\Sigma$ in terms of the potential parameters given in Eq.~(\ref{eq:fsigma}),
the scalar mass ratio becomes
\be
\frac{M_{\phi_I}}{M_\Theta}  =   \sqrt{ 3\lambda + \kappa} \left(  2\lambda + \kappa + \frac{\kappa}{ 3}  \, \sqrt{4(3 \lambda + \kappa) \, \frac{ m^2_\Sigma  }{ \mu_\Sigma^2}   + 1}  \; \right)^{\! -1/2}         ~~.
\label{eq:massRatio}
\ee

The condition (\ref{eq:bounded}) implies $2 \lambda +  \kappa > 0$, so that the maximum value of $M_{\phi_I}/M_\Theta$ occurs at $\kappa < 0$ and $m_\Sigma^2 > 0$.
Furthermore, $m_\Sigma^2/\mu_\Sigma^2$ must be maximized to reach the maximum for $M_{\phi_I}/M_\Theta$.
The vacuum that preserves $SU(3)_c$ is the global minimum for $\kappa < 0$ (which implies $\lambda >0$) and $m_\Sigma^2 > 0$ 
provided \cite{Bai:2017zhj}
\be 
(3 \lambda + \kappa)  \, \frac{m_\Sigma^2}{\mu_\Sigma^2}   \, \leq \,   {\cal G}(\kappa/\lambda)  ~~,
\label{eq:221line}
\ee
where the function $ {\cal G}(x)$ is defined on the interval  $-1 < x < 0$ by
\be
 {\cal G}(x) =  (3 + x) \left( \frac{ \left( 4  + 2 x \right)^{3/2} \!\! \!}{ \sqrt{ 1+ x}} \, -2(4  + x)  \right)^{\! -1}  ~~.
\label{eq:xi}
\ee
Note that if the inequality (\ref{eq:221line}) is not satisfied, then the gauge symmetry breaking pattern in the global minimum is 
$SU(3)_1\times SU(3)_2 \to SU(2)\times SU(2) \times U(1)$, so color is broken. Nevertheless, an $SU(3)_c$-symmetric local 
minimum still exists  \cite{Bai:2017zhj} for slightly larger values of $m_\Sigma^2/\mu_\Sigma^2$. We will not consider here the possibility that the 
viable vacuum is not the global minimum. 

The constraint (\ref{eq:221line}) implies the following upper limit:
\be
 \frac{M_{\phi_I}} {M_\Theta} \,  \leq \,  R_{\rm max} (\kappa/\lambda) ~~,
 \label{eq:massRatio1}
\ee 
 where the function introduced here is
\be
R_{\rm max} (x) = \frac{\sqrt{3 + x}}{  \left( 2 + x + (x/3) \, \sqrt{ 4  {\cal G}(x)    + 1} \right)^{\! 1/2} }   ~~~.
\ee
For any values of the quartic couplings that satisfy $-1 < \kappa/\lambda < 0$ we find that the upper limit for the  mass ratio is above $\sqrt{3}$.
The maximum value $(M_{\phi_I}/M_\Theta)_{\rm max}  =  3/\sqrt{2}$ is reached at $\kappa \to 0$. Hence,
the ReCoM predicts 
\be
M_{\phi_I} \lesssim 2.1  M_\Theta   ~~.
\ee

\begin{figure}[t]
\begin{center}
\vspace*{-0.1cm}
\includegraphics[width=0.66\textwidth]{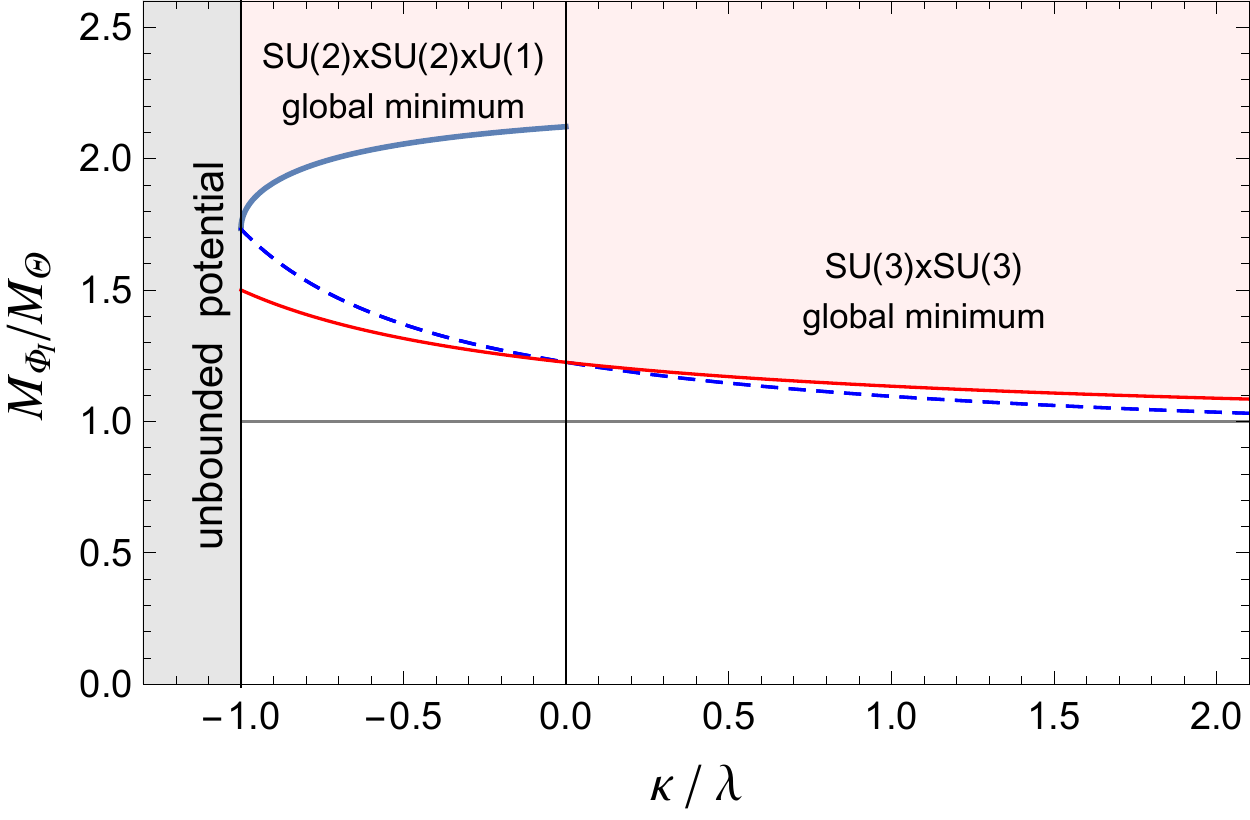}
\vspace*{-0.3cm}
\caption{Allowed range (unshaded region) for 
the scalar mass ratio $M_{\phi_I}/M_\Theta$ as a function of the quartic coupling ratio $\kappa/\lambda$.
In the pink shaded region the global minimum does not preserve color (for $\kappa/\lambda < 0$) or the extended gauge symmetry 
is unbroken (for $\kappa/\lambda > 0$). In the gray shaded region the potential is not bounded from below. The upper limit 
on $M_{\phi_I}/M_\Theta$ is given by the solid blue line for $m_\Sigma^2 > 0$ and  $\kappa < 0$; the solid red line for $m_\Sigma^2 > 0$ and $\kappa < 0$; the dashed blue line for $\kappa\,m_\Sigma^2 > 0$.
} 
\label{fig:massRatio}
\end{center}
\end{figure}

Let us now find the maximum values of $M_{\phi_I}/M_\Theta$ in other regions of parameter space.
For $\kappa \ge 0$ and $m_\Sigma^2 >  0$ the only vacuum is $SU(3)_c$ symmetric. From Eq.~(\ref{eq:massRatio}),
the maximum of $M_{\phi_I}/M_\Theta$  is obtained when $m_\Sigma^2 \to 0$, so that 
\be
\frac{M_{\phi_I}}{M_\Theta}  \leq  \sqrt{ \frac{3\lambda + \kappa}{2\lambda + 4\kappa/3}}        ~~~.
\label{eq:massRatio2}
\ee
The range of  $M_{\phi_I}/M_\Theta$ consistent with a viable global minimum is shown as a function of $\kappa/\lambda$ 
in Figure~\ref{fig:massRatio}. The solid blue line there is the upper limit for $\kappa < 0$, given in (\ref{eq:massRatio1}), while
the dashed blue line represents the upper limit  (\ref{eq:massRatio2}) for $\kappa > 0$.

For  $m_\Sigma^2 < 0$, the $SU(3)_c$ symmetric vacuum is the global minimum provided \cite{Bai:2017zhj}
\be
(3 \lambda + \kappa) \frac{m_\Sigma^2}{\mu_\Sigma^2} > -\frac{2}{9}   ~~~.
\label{eq:globalNeg}
\ee
If this is not satisfied, the $SU(3)_1\times SU(3)_2$ gauge symmetry is unbroken. 
If $\kappa < 0$, then the maximum is at $m_\Sigma^2 \to 0$ and given by \eqref{eq:massRatio2}.
If $\kappa > 0$ and  $m_\Sigma^2 < 0$, then $M_{\phi_I}/M_\Theta$ reaches its maximum when the above inequality is saturated:
\be
\frac{M_{\phi_I}}{M_\Theta}  \leq  \sqrt{ \frac{3\lambda + \kappa}{2\lambda + 10\kappa / 9}}     ~~~.
\ee
The above upper limit is shown as the solid red lines in Figure~\ref{fig:massRatio}. 
The conclusion is that even though it is natural to have  $M_{\phi_I} < M_\Theta$,  the opposite mass relation,
$M_{\phi_I} > M_\Theta$, also occurs for sizable regions of parameter space.

For the other singlet scalar, $\phi_R$,  the squared mass given  in Eq.~(\ref{eq:scalar-masses}) may be rewritten 
using the expression of the $\Sigma$ VEV, $f_\Sigma$, from \eqref{eq:fsigma}:
\bear
M^2_{\phi_R} = \frac{f_\Sigma}{\sqrt{6} } \sqrt{ 4( 3\lambda + \kappa ) m_\Sigma^2 + \mu_\Sigma^2} ~~~.
\eear
For $m^2_\Sigma < 0$, the condition (\ref{eq:globalNeg}) implies 
\be
\frac{1}{3} < \frac{M_{\phi_R}}{M_{\phi_I}}  < \frac{1}{\sqrt{3}}  ~~. 
\ee
For $m^2_\Sigma > 0$,  the condition \eqref{eq:bounded} for the potential to be bounded from below implies a lower  limit on $M_{\phi_R}/M_{\phi_I}$:
\be
\frac{M_{\phi_R}}{M_{\phi_I}} > \frac{1}{\sqrt{3}}  ~~.
\ee
Note that ${\phi_R}$ is much heavier than other scalars when $\lambda \gg |\kappa|$ and $m^2_\Sigma \gg \mu^2_\Sigma$. 

\bigskip

\section*{Appendix B: Coloron decays into scalars}
\renewcommand{\theequation}{B.\arabic{equation}}
\addcontentsline{toc}{section}{\normalsize Appendix B: Coloron decays into scalars}
 \setcounter{equation}{0}
\label{app:coloron-decays}

A simple way to compute the partial width of the coloron decay into a pair of color-octet scalars is by fixing the color index of the coloron. Let us compute the width of $G^{\prime 8}$.
$SU(3)_c$  gauge invariance then ensures that $G^{\prime a}$  has the same width for any $a = 1,...,8$. The interactions of $G^{\prime 8}$ with a pair of color-octet scalars,
included in Eq.~(\ref{eq:GpScalars}), is 
\be
- \sqrt{3}  \, g_s \,  \frac{ 1 - \tan^2\!\theta }{4 \tan \theta} \; G_\mu^{\prime\,8} 
\, \left(  \Theta^4 \, \partial^\mu \Theta^5  -  \Theta^5 \, \partial^\mu \Theta^4  +   \Theta^6 \, \partial^\mu \Theta^7  -  \Theta^7 \, \partial^\mu\Theta^6 \right)   ~~.
\label{eq:GpScalars8}
\ee
Here we took into account that the only nonzero values (up to index permutations) of the  totally-antisymmetric color tensor $f^{abc} $ with one index being 8 are 
$f^{458} = f^{678} = \sqrt{3}/2$. 

The matrix element for $G^{\prime 8} \to \Theta^4 \Theta^5$ is 
\be
{\cal M} \left(  G^{\prime 8} \to \Theta^4 \Theta^5  \right) =  i \sqrt{3}  \, g_s \,  \frac{ 1 - \tan^2\!\theta }{4 \tan \theta} \;  (p_5^\mu - p_4^\mu) \;  \varepsilon_\mu (P)  ~~,
\ee
where  $\varepsilon_\mu$ is the polarization vector of  the coloron,  and   $p_4^\mu$,  $p_5^\mu$, $P^\mu$ are the 4-momenta of  $\Theta^4$, $ \Theta^5$, $G^{\prime 8} $, respectively.
The matrix element for $G^{\prime 8} \to \Theta^6 \Theta^7$ has the same form. The two processes do not interfere, so that 
the squared matrix element  for coloron decays into a pair of 
$\Theta$ scalars, averaged over the coloron polarizations and summed over the two possible final states ($\Theta^4 \Theta^5$ and  $\Theta^6 \Theta^7$),  is given by
\be
\left| \overline {\cal M} \left(  G^{\prime} \to \Theta\Theta  \right)  \right|^2 = g_s^2 \,  \frac{ (1 - \tan^2\!\theta)^2 }{8 \tan^2\!\theta} \, \left( M_{G^\prime}^2 - 4 M_\Theta^2 \right)  ~~.
\label{eq:element-squared}
\ee
This implies that the partial width of the coloron decaying into two $\Theta$'s is the one given in Eq.~(\ref{eq:Gwidths}).
That partial width corrects a factor of 2 missing from the one given in Ref.~\cite{Bai:2010dj}. 
The result in Eq.~(\ref{eq:element-squared}) can also be checked by using the interaction in Eq.~(\ref{eq:GpScalars}),
and summing over all $a,b,c$ indices; when squaring the matrix element, it is necessary to notice  that 
$f^{abc} \Theta^b \, \partial^\mu \Theta^c$ can be contracted with both $f^{abc} \Theta^b \, \partial^\mu \Theta^c$  and $f^{acb} \Theta^c \, \partial^\mu \Theta^b$, which again gives a factor of 2. 
 
The averaged  squared matrix element  for the asymmetric  decay $G' \to \Theta \, \phi_I$ is slightly more complicated because the scalar masses are different:
\be
\left| \overline {\cal M} \left(  G^{\prime} \!\to \Theta \, \phi_I \right)  \right|^2 = g_s^2 \,  \frac{ (1 + \tan^2\!\theta)^2 }{18\tan^2\!\theta \; M_{G^\prime}^2}  
\left[ M_{G^\prime}^2 \!-\! \left( M_\Theta \!+\! M_{\phi_I}\right)^2 \right]  \left[ M_{G^\prime}^2  \!-\! \left( M_\Theta \!-\! M_{\phi_I}\right)^2 \right]   ~.
\ee
The corresponding partial width for $G' \to \Theta \, \phi_I$  is given in Eq.~(\ref{eq:Gwidths}), and  agrees with the result of Ref.~\cite{Bai:2010dj}.

\bigskip 

\providecommand{\href}[2]{#2}\begingroup\raggedright\endgroup

\end{document}